\documentclass[twocolumn,english,prb,notitlepage,superscriptaddress,nobibnotes]{revtex4-2}
\usepackage[T1]{fontenc}
\usepackage[latin9]{inputenc}
\usepackage{graphicx}
\usepackage{amsmath}
\usepackage{amssymb}
\usepackage{float}
\usepackage[colorlinks = true,
linkcolor = blue,
urlcolor  = blue,
citecolor = blue,
anchorcolor = blue]{hyperref}
\usepackage[usenames, dvipsnames]{color}

\makeatletter
\usepackage{babel}
\usepackage{float}
\usepackage{appendix}

\usepackage{tabularx}
\newcolumntype{C}{>{\centering\arraybackslash}X}

\makeatother

\begin{document}

\title{Nonlinear light cone spreading of correlations in a triangular quantum magnet: a hard quantum simulation target}

\author{A. Scheie} 
\email{scheie@lanl.gov}
\affiliation{MPA-Q, Los Alamos National Laboratory, Los Alamos, NM 87545, USA}

\author{J. Willsher}  
\affiliation{Technical University of Munich, TUM School of Natural Sciences, Physics Department, 85748 Garching, Germany}
\affiliation{Munich Center for Quantum Science and Technology (MCQST), Schellingstr. 4, 80799 M\"unchen, Germany}
\affiliation{Max-Planck-Institut f\"ur Physik komplexer Systeme, 01187 Dresden, Germany}

\author{E. A. Ghioldi} 
\affiliation{Theoretical Division, Los Alamos National Laboratory, Los Alamos, NM 87545, USA}
\affiliation{Institute for Material Science, Los Alamos National Laboratory, Los Alamos, NM 87545, USA}
\affiliation{Department of Physics and Astronomy, University of Tennessee, Knoxville, Tennessee 37996, USA} 

\author{Kevin Wang}   
\affiliation{Department of Physics, University of California, Berkeley, CA 94720, USA}

\author{P. Laurell}   
\affiliation{Department of Physics and Astronomy, University of Missouri, Columbia, Missouri 65211, USA} 
\affiliation{Materials Science and Engineering Institute, University of Missouri, Columbia, Missouri 65211, USA} 

\author{J. E. Moore} 
\affiliation{Department of Physics, University of California, Berkeley, CA 94720, USA}

\author{C. D. Batista}  
\affiliation{Department of Physics and Astronomy, University of Tennessee, Knoxville, TN 37996, USA}
\affiliation{Shull Wollan Center, Oak Ridge National Laboratory, TN 37831. USA}

\author{J. Knolle}   
\affiliation{Technical University of Munich, TUM School of Natural Sciences, Physics Department, 85748 Garching, Germany}
\affiliation{Munich Center for Quantum Science and Technology (MCQST), Schellingstr. 4, 80799 M\"unchen, Germany}

\author{D. Alan Tennant}
\affiliation{Department of Physics and Astronomy, University of Tennessee, Knoxville, Tennessee 37996, USA}

\date{\today}

\begin{abstract}

Dynamical correlations of quantum many-body systems are typically analyzed in the momentum space and frequency basis. However, quantum simulators operate more naturally in real space, real time settings. Here we analyze the real-space time-dependent van Hove spin correlations $G(r,t)$ of the 2D triangular antiferromagnet KYbSe$_2$ as obtained from high-resolution Fourier-transformed neutron spectroscopy. We compare this to  $G(r,t)$ from five theoretical simulations of the well-established spin Hamiltonian. Our analysis reveals non-linear sub-ballistic low-temperature transport in KYbSe$_2$ which none of the current state-of-the-art numerical or field-theoretical methods reproduce. Our observation signals an emergent collective hydrodynamics, perhaps associated with the quantum critical phase of a quantum spin liquid, and provides an ideal benchmark for future quantum simulations.  

\end{abstract}

\maketitle

\section{Introduction}

Often in science, we know a system's governing equations without understanding its collective behavior. 
Quantum phases of matter are a prime example: the interactions between atoms are often well-understood, but the many-body collective effects are too complex to compute \cite{Anderson-MoreIsDifferent,Laughlin_Nobel_1999}.  
Such ``emergent behaviors'' are a long-standing focus of condensed matter physics; but they are also valuable benchmarks for quantum simulations \cite{bartschi2024potential,ALEXEEV2024666}, especially for regions without simple quasiparticle descriptions. If a quantum simulation (either from qubit hardware or classical high-performance computing techniques) can match the experimental many-body physics, this validates the simulation as correct. The requirements are (i) a system whose underlying Hamiltonian is known precisely, (ii) a clear observable from high-quality experimental data, and (iii) the failure of conventional simulation techniques. 
Quantum spin systems provide an excellent platform for such benchmarks, as the interacting spins can be directly mapped to coupled qubit networks. 
In this context we present an experimental study of the collective spin dynamics of the two-dimensional (2D) triangular magnet KYbSe$_2$.

\begin{figure}
 	\includegraphics[width=\columnwidth]{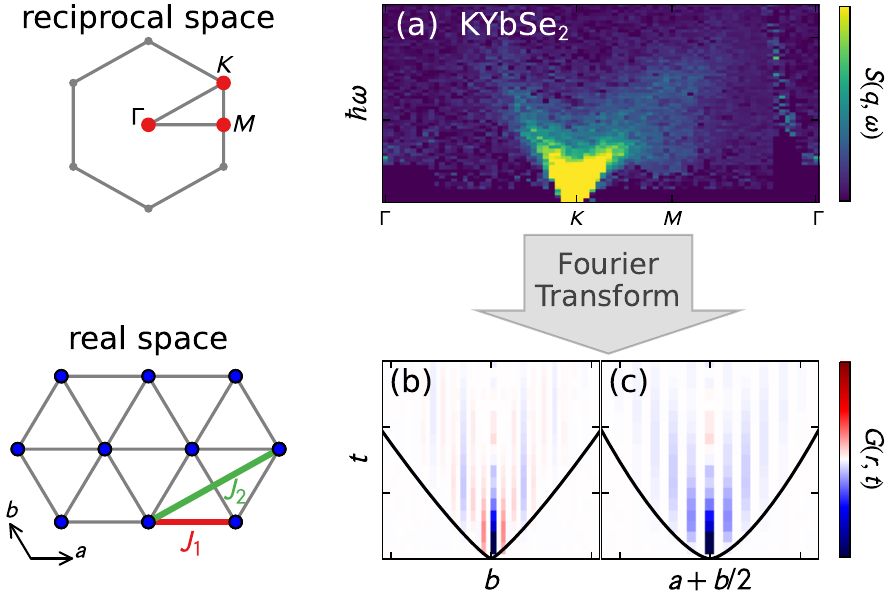} 
 	\caption{Schematic of generating triangular-lattice real-space correlations. $S(q,\omega)$, which gives spin correlations in reciprocal space and frequency, is Fourier transformed into $G(r,t)$, which gives spin correlations in real-space and time.}
 	\label{fig:schematic}
 \end{figure}

 KYbSe$_2$ is a material with a 2D triangular lattice of magnetic Yb$^{3+}$ ions \cite{Xing_2021_KYS,Scheie2024_KYS}  (Fig. \ref{fig:schematic}). Its spin interactions are nearly perfect Heisenberg (isotropic) exchanges \cite{Scheie2024_KYSNYS}, and it lies on a well-studied theoretical $J_2/J_1$ phase diagram with $J_2/J_1=0.044(5)$ \cite{Scheie2024_KYSNYS}. Despite the presence of weak magnetic order below $T_N=290$~mK, the inelastic neutron spectrum is dominated by a diffuse spectrum with a well-defined lower bound, signaling a breakdown in conventional magnon physics \cite{Scheie2024_KYS}.

 The 2D triangular Heisenberg antiferromagnet is also a well-studied but unsolved theoretical problem. Numerical simulations show a quantum spin liquid (QSL) phase when $J_2/J_1$ exceeds $\approx 0.06$~\cite{PhysRevB.92.041105,PhysRevB.92.140403,PhysRevB.93.144411,PhysRevB.94.121111,PhysRevB.95.035141,PhysRevB.96.075116,PhysRevLett.123.207203,Gallegos2025}, but field-theoretic methods disagree on the predicted QSL neutron spectra \cite{Scheie2024_KYS,Willsher_2025} and matrix product methods are limited by finite-size effects in time and space \cite{PhysRevLett.129.227201,Sherman_2023_Spectral,Drescher_2023,drescher2025spectral}. Thus the 2D triangular antiferromagnet meets all three criteria (listed above) for a quantum benchmark problem. 
 
In this study, we analyze the two-dimensional (2D) spin correlations of triangular KYbSe$_2$ \cite{Xing_2021_KYS,Scheie2024_KYS} in real space and time, and compare to simulations of the 2D triangular Heisenberg antiferromagnet (Fig. \ref{fig:schematic}). 
We analyze the real-space dynamics for two reasons: (1) it has revealed nontrivial information about quantum dynamics in one dimension \cite{Scheie2022,scheie2024reconstructing,kish2024high} and (2) it is a natural output of quantum computers \cite{kumaran2025quantum,lee2025digital,rosenberg2024dynamics,Liang_2025,Manovitz2025,chowdhury2025quantum,haghshenas2025digital}. 
We find an experimental nonlinear light cone at $T=0.3$~K that all our theoretical methods fail to capture. This implies a collective behavior that is both a target for quantum simulations, and a frontier to understand in quantum condensed matter physics.


\begin{figure*}
    \includegraphics[width=\textwidth]{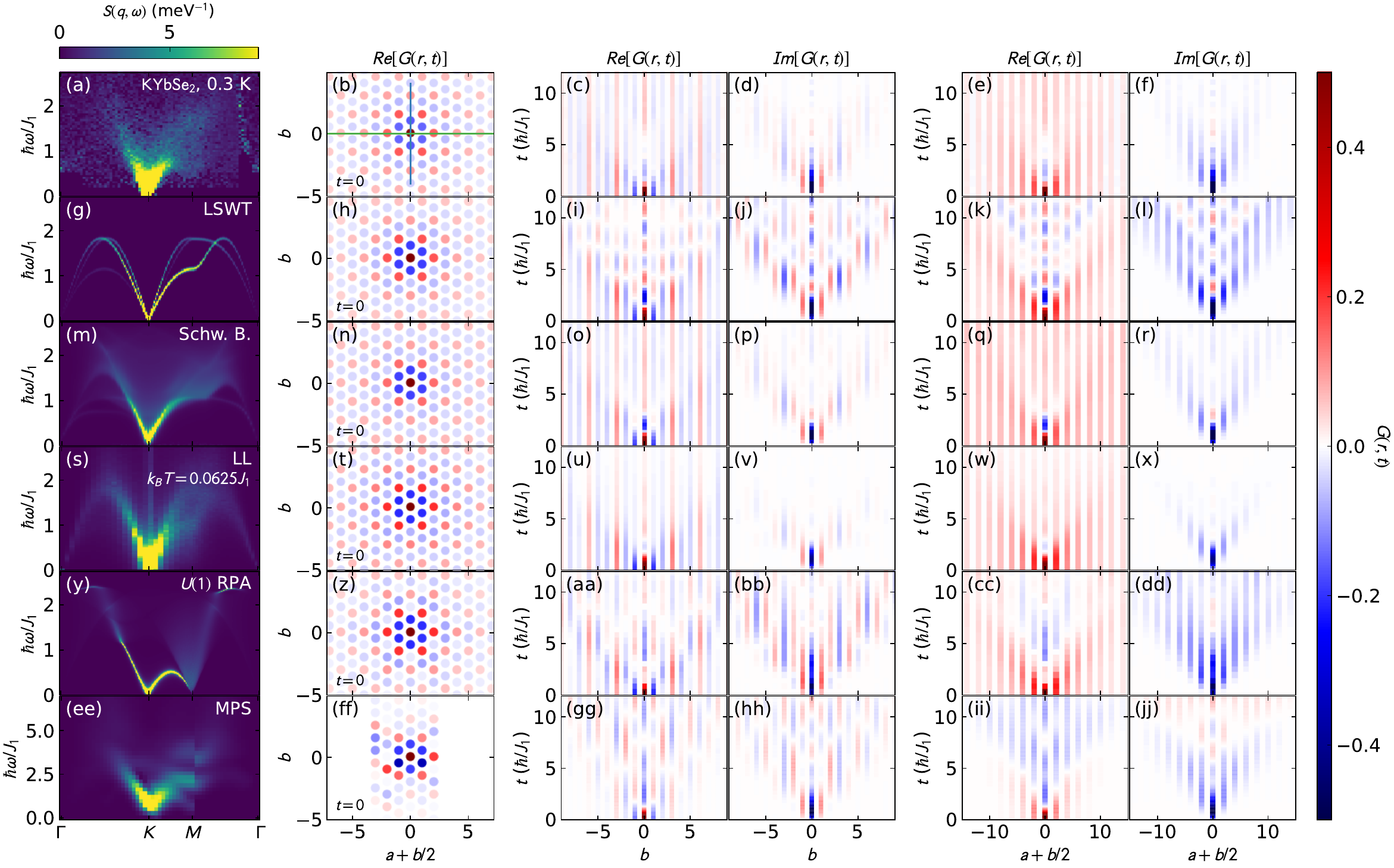} 
    \caption{Van Hove correlations for KYbSe$_2$ (a)-(f) and various theoretical models: linear spin wave theory (LSWT) (g)-(l), Schwinger bosons (m)-(r), finite-temperature Landau-Lifshitz dynamics (LL) (s)-(x), $U(1)$ random phase approximation (RPA) simulations (y)-(dd), and matrix product state (MPS) calculations on a six-site circumference cylinder (ee)-(jj). 
    The left column shows the spectra along high-symmetry directions, the second column shows the $t=0$ spin correlation pattern (red indicates ferromagnetic spin correlation, blue indicates antiferromagnetic spin correlation). The third and fourth columns show the real and imaginary spin correlations along the first neighbor crystallographic direction $b$ [vertical blue line in panel (b)]. Columns five and six show the real and imaginary spin correlations along the second neighbor crystallographic direction ${\bf a+b}/2$ [horizontal green line in panel (b)].
    Light cones are clearly visible in all cases, but poor agreement is found with LSWT whereas much better agreement is found with the Schwinger Boson and LL calculations.
    }
    \label{fig:ModelComparison}
\end{figure*}

\section{Results}

We obtain the real-space magnetic Van Hove correlation function $G(r,t)$ from KYbSe$_2$ inelastic neutron scattering $S(q,\omega)$ at $T=0.3$~K as described in the Methods section. 
We also calculate the 2D $G(r,t)$ from five different theoretical descriptions (using the KYbSe$_2$ Hamiltonian of Ref. \cite{Scheie2024_KYSNYS}): linear spin wave theory (LSWT) at $T=0$ as calculated with the \textit{SpinW} software \cite{SpinW}, Schwinger boson theory at $T=0$\cite{Scheie2024_KYS,Ghioldi_2018}, classical Landau Lifshitz (LL) dynamics at finite temperature calculated using \textit{SU(N)NY} software \cite{SUNNY_Zhang_2021,SUNNY_Lane_2024,dahlbom2025sunnyjljuliapackagespin}, random phase approximation (RPA) simulations at $T=0$ from a $U(1)$ quantum spin liquid \cite{Willsher_2025}, and Matrix Product Simulations (MPS) at $T=0$\cite{Sherman_2023_Spectral}. 
Each of these five simulations are plotted in Fig. \ref{fig:ModelComparison}, and each assumes different ground states and dynamics. LSWT treats the system as well-ordered with magnon quasiparticle excitations without lifetime or decay effects, Schwinger bosons treat the excitations as a bound state of two spinons \cite{Ghioldi_2018,Ghioldi22,Dey_2024_Field}, LL treats the spins as classical vectors (though with renormalized intensity to enforce the quantum sum rule \cite{Huberman_2008,Dahlbom_2024_Quantum}) thus neglecting multi-spin quantum effects, 
RPA calculations assume interacting spinons of a $U(1)$ Dirac spin liquid that is here tuned to its phase transition to $120^{\circ}$ magnetic order \cite{Willsher_2025}, 
and MPS obtained an approximate wave function to the many-body spin system on a 6-site circumference cylinder. 
This diversity of theoretical models allows us to compare the $G(r,t)$ signatures of qualitatively different types of excitations to each other and to experimentally measured  KYbSe$_2$ spin dynamics.

At $T=0.3$~K, the KYbSe$_2$ correlation function  $G(r,t)$ shown  in Fig. \ref{fig:ModelComparison} exhibits a clear ``light cone'' in both the real and imaginary components (by analogy with special relativity, the light cone separates the static and dynamic regions and is defined by the fastest moving quasiparticles \cite{Scheie2022}). Interestingly, the onset of the light cone is non-uniform, with the third neighbor along $b$ showing evolving correlations (nonzero ${Im}[G(r,t)]$) before the second neighbor along $b$. In the real part at $T=0.3$~K [Fig. \ref{fig:ModelComparison}(c) and (e)], KYbSe$_2$ shows a temporary inversion of $r=0$ spin correlations  but at all finite distances ($|r|>0$) merely a modulation of the spin correlation pattern. At elevated temperatures $T=1$~K and  $T=2$~K  (shown in the Supplemental Information \cite{SuppMat}), the spin pattern is fully inverted above the light cone, leaving a spin pattern opposite the initial state. This oscillation is reminiscent of what was called a ``Higgs-mode'' in cold-atom simulations of quantum criticality \cite{Manovitz2025}.  

In addition to the two-point spin correlator $G(r,t)$, we also compute the Quantum Fisher Information Matrix (QFIM) of KYbSe$_2$ \cite{scheie2024reconstructing} (shown in the Supplemental Information \cite{SuppMat}). This quantity describes the quantum correlations as a function of distance in the lattice. Conveniently, it is directly related to the $T=0$ variance \cite{Hauke2016} and is straightforwardly calculable with quantum computers \cite{meyer2021fisher}, making it useful as an additional benchmark.  

\subsection*{Nonlinear light cones}

\begin{figure*}
    \includegraphics[width=\textwidth]{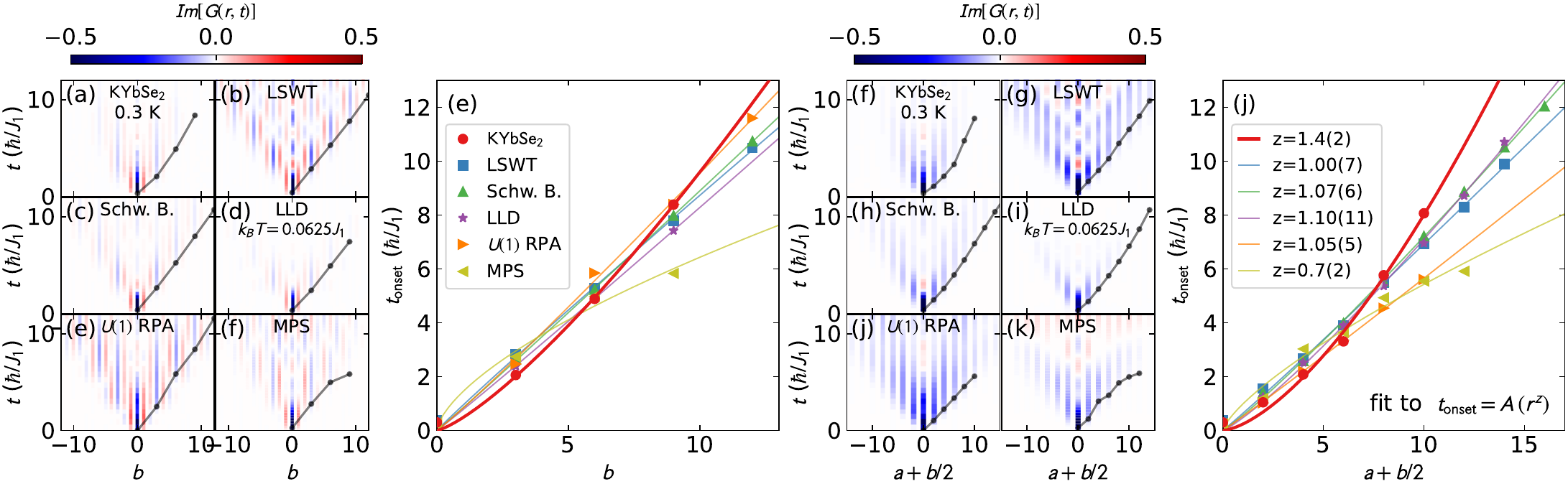} 
    \caption{Sub-ballistic transport in KYbSe$_2$. Panels (a)-(d) and (f)-(i) show the light cones of KYbSe$_2$ alongside various theoretical models, where the black points show the onset (defined as the midpoint of the rist in $G(r,t)$). Panels (e) and (j) plot the onset time versus distance. The solid lines are fitted to $t_{\rm onset} = A(r^z)$ where $A$ and $z$ are fitted constants (both directions fitted simultaneously). The fitted exponents are listed in panel (j), with the experimental KYbSe$_2$ noticeably deviating from the theoretical simulations. Instead of being in a ballistic regime ($z \approx 1$), the KYbSe$_2$ light cone exhibits sub-ballistic superdiffusive behavior ($z \approx 1.5$). 
    }
    \label{fig:Superdiffusivity}
\end{figure*}

The different theoretical models capture various aspects of the experimental $G(r,t)$, detailed in the Supplemental Information \cite{SuppMat}. 
However, there is one clear feature revealed by 2D $G(r,t)$ that none of the theoretical simulations predict: a nonlinear light cone. 
In Fig. \ref{fig:Superdiffusivity} we plot the imaginary $G(r,t)$ for KYbSe$_2$ and the five theoretical simulations, picking out the onset of the light cone as the midpoint of the rise in ${Im}[G(r,t)]$. 

All the theoretical simulations show an approximately linear light cone (except MPS which has a downward bend beyond six lattice units because of finite size effects). 
KYbSe$_2$, meanwhile, shows a marked upward bend at around the six lattice units, both along the near neighbor and next-near-neighbor directions. (In the Supplemental Information \cite{SuppMat} we show this is not an effect of experimental resolution broadening.) We quantify this nonlinearity by fitting a simple equation $t_{\rm onset} = A(r^z)$ along the $b$ and $a+b/2$ directions simultaneously.
The results are shown in Fig. \ref{fig:Superdiffusivity}(e) and (j), revealing that KYbSe$_2$ has a dynamical exponent $z=1.4(2)$.  For ballistic transport, one expects a linear light cone with $z=1$; for diffusive transport, one expects a quadratic light cone with $z=2$ \cite{DeNardis_2020,Scheie2021KPZ}. KYbSe$_2$ has a superdiffusive dynamical exponent that is between these limits.

\section{Discussion}

The nonlinear KYbSe$_2$ light cone reveals some fundamental collective behaviors, likely related to the proximate quantum spin liquid phase, that current simulation methods do not capture. If we assume that the observed dynamics can be described by a single asymptotic scaling regime, i.e., that the data do not reflect a crossover between distinct transport behaviors, the resulting light cone is consistent with subdiffusive scaling. 
What causes this subdiffusivity? The data and simulations here are insufficient to say for certain, but we consider three hypotheses.

The first hypothesis is superdiffusive dynamics seen in integrable 1D spin chains \cite{PhysRevLett.106.220601,RevModPhys.93.025003,Bulchandani_2021,Scheie2021KPZ,kish2024high,rosenberg2024dynamics,joshi2022observing} where the dynamical exponent $z=1.5$ has been observed as a low-frequency intensity power law in KCuF$_3$ \cite{Scheie2021KPZ}, and a nonlinear $G(r,t)$  light-cone in YbAlO$_3$ \cite{kish2024high}. In cold-atom experiments, this was observed as a short-time oscillation evolving at long-times to emergent hydrodynamics witnessed by a nonlinear spread of spin correlations \cite{joshi2022observing}, very similar to KYbSe$_2$. 
However, the 2D triangular lattice is non-integrable and therefore not subject to the same conservation constraints (potentially rendering its transport diffusive \cite{Bohrdt_2017}).  Furthermore, in 1D the superdiffusivity was manifest at the highest temperatures \cite{Scheie2021KPZ,kish2024high,joshi2022observing}, where as in KYbSe$_2$ this effect occurs at the lowest temperatures. Therefore, the same integrable high-temperature physics does not apply in KYbSe$_2$. 

A second hypothesis is that the nonlinear light cone might be associated with confined spinon quasiparticles. KYbSe$_2$ is known to order magnetically, such that at short times the spinons are free but at longer time scales are confined and propagate slowly, again as observed in simple spin chains of domain-wall confinement \cite{kormos2017real,Vovrosh2021}. However, the timescales do not match. If the timescale of such coalescing quasiparticles is on the energy scale $T_N= 0.29$~K \cite{Scheie2024_KYS}, this would correspond to $\frac{h}{k_B T_N} = 160$~ps, or 110 in units of $\hbar/J_1$. This is an order of magnitude longer timescale than the observed nonlinearity, which appears at $t \approx 4 (\hbar/J_1)$. Thus the observed nonlinear light cone does not appear to be from spinon confinement. 

A third hypothesis to explain low-temperature nonlinearity is emergent quantum critical hydrodynamics. In the limit of strong interactions (e.g. at quantum critical points), the quasiparticle picture breaks down in favor of a hydrodynamic description \cite{hartnoll2018holographic}, which has been remarkably successful in describing one-dimensional quantum matter \cite{PhysRevX.15.010501}. 
The nonlinear light cone may be a signature of low-temperature emergent hydrodynamics, given the observed critical scaling in KYbSe$_2$~\cite{Scheie2024_KYS}. 
This is supported by theoretical proposals of quantum hydrodynamics near QSL confinement~\cite{PhysRevB.104.235412}, and the recent suggestion that the low-energy response of this material may be described by Gross-Neveu-QED$_3$ universality with quasiparticle breakdown \cite{Willsher_2025}. 
Similarly, the nonlinearity it could signal a crossover from a short-time ballistic regime to an intermediate diffusive regime associated with the critical fluctuations, noting that our experiment temperature is just above $T_N=0.29$~K. 
Finally, we study semiclassical disorder in the Supplemental Information using LL, and show that it is not capable of describing a non-linearity of this magnitude.

Whatever the explanation, the $G(r,t)$ analysis reveals a collective sub-ballistic behavior in KYbSe$_2$ that none of the model calculations reproduce.  Whether this is a feature of the spin liquid phase, the quantum critical regime, or something else entirely, will require further study. 

In addition to being an intriguing fundamental science question, the nonlinear dynamic exponent is important because it meets the criteria for an ideal quantum benchmark problem: (i) the underlying KYbSe$_2$ Hamiltonian is known to high precision \cite{Scheie2024_KYSNYS}, (ii) clean experimental data exist  with a clear experimental signature to target (nonlinearity in the light cone of a proximate spin liquid) and (iii) state-of-the-art HPC methods are insufficient to describe this behavior. 
With the rapid advances in digital \cite{lee2025digital,haghshenas2025digital} and analog \cite{rosenberg2024dynamics,Bauer_2025_Progress} quantum simulation of spin dynamics, matching this feature is a realistic near-term goal for quantum computers.  
Thus this signal is an opportunity to advance and validate quantum technology while also addressing a fundamental question in many-body physics: how do quantum interactions change how correlations spread through a lattice? 


\section{Conclusion}

We have analyzed the $G(r,t)$ of 2D triangular KYbSe$_2$ and compared it with several theoretical simulations of the same spin interaction model. This comparison revealed a nonlinear light cone that was unanticipated by previous work. 

The key result of sub-ballistic magnetic transport will require some theoretical work to understand, but it signals a  deficiency in our theoretical modeling approaches. It also makes an excellent benchmark to target for quantum computation \cite{bartschi2024potential} or HPC quantum simulation methods \cite{pugzlys2025autoregressiveneuralnetworkextrapolation}. 
Finally, this study highlights the utility of $G(r,t)$ in 2D as a way to study collective dynamics of quantum systems---even those that are extremely challenging to simulate. Some features which are very subtle in reciprocal space become clear in real-space. 


\section*{acknowledgments}

 The work by A.S., E.A.G., J.E.M., C.D.B., and D.A.T. was primarily supported by the Quantum Science Center (QSC), a National Quantum Information Science Research Center of the U.S. Department of Energy (DOE). 
 Parts of the work by A.S. were performed at Aspen Center for Physics, which is supported by National Science Foundation grant PHY-2210452. 
 Final stages of analysis and writing by A.S. were supported by the Laboratory Directed Research and Development program of Los Alamos National Laboratory under project number 20250836ECR. E.A.G. also acknowledges funding from the LANL LDRD program for final stages of analysis. 
K.W. and J.E.M. used resources provided by the National Energy Research Scientific Computing Center, a DOE Office of Science User Facility supported by the Office of Science of the U.S. Department of Energy under Contract No. DE-AC02-05CH11231 using NERSC award BES-ERCAP0032440. K.W. was supported through a fellowship from the Kavli Energy NanoScience Institute (ENSI).
J. K. and J. W. acknowledge support from the Deutsche Forschungsgemeinschaft (DFG, German Research Foundation) grant TRR 360 - 492547816 [14]. J.K. further acknowledgess support from DFG under Germany's Excellence Strategy (EXC-2111-390814868), DFG Grants No. KN1254/1-2, KN1254/2-1  and SFB 1143 (project-id 247310070), 
as well as the Munich Quantum Valley, which is supported by the Bavarian state government with funds from the Hightech Agenda Bayern Plus. 
J.K. further acknowledges support from the Imperial-TUM flagship partnership, as well as the Keck Foundation. 
 We acknowledge helpful discussions with Gabor Halasz and Michael Knap. 

\section*{Methods}\label{sec:Methods}

Magnetic inelastic neutron scattering can be analyzed by taking a Fourier transform in space and time from the structure factor $S(q,\omega)$ to the Van Hove correlation function $G(r,t)$ \cite{PhysRev.95.1374,Scheie2022,PhysRevB.106.085110}. This analysis has been performed for 1D spin chains KCuF$_3$ \cite{Scheie2022,scheie2024reconstructing} and YbAlO$_3$ \cite{kish2024high}. 
Here we extend this analysis beyond 1D to the 2D triangular magnet KYbSe$_2$.

We analyze the KYbSe$_2$ magnetic scattering data reported in Ref. \cite{Scheie2024_KYS} at $T=0.3$~K, 1~K, and 2~K. Data were corrected for the isotropic Yb$^{3+}$ form factor \cite{BrownFF}, and the 2D scattering volume between (-1,-1,0) and (1,1,0) was filled out by folding further Brillouin zones into the first Brillouin zone and symmetrizing. Carrying out the three-dimensional (3D) Fourier transform (two spatial dimensions and time), we obtain the $G(r,t)$ signals shown in Fig. \ref{fig:ModelComparison}(a)-(f) at 0.3~K and in the Supplementary Information \cite{SuppMat} at elevated temperatures.  
In all figures in this study, we plot two orthogonal cuts through ${\bf r}=(0,0)$: along the nearest neighbor direction ($\bf b$) and along the second neighbor direction (${\bf a+b}/2$).  

KYbSe$_2$ is a highly isotropic system, which made the analysis $G(r,t)$ very straightforward. More anisotropic spin systems will require considering the polarization factor. Nevertheless, with the proper measurements and modeling,  $G(r,t)$ can in principle be analyzed for any magnetic signal.

%


\newpage

\renewcommand\thefigure{S.\arabic{figure}}
\setcounter{figure}{0}
\renewcommand{\theequation}{S\arabic{equation}}
\setcounter{equation}{0}
\renewcommand{\thetable}{S\Roman{table}}
\setcounter{table}{0}

\section*{Supplemental Information for Nonlinear light cone spreading of correlations in a triangular quantum magnet: a hard quantum simulation target}

\section{Time evolution}

\begin{figure*}
    \includegraphics[width=0.85\textwidth]{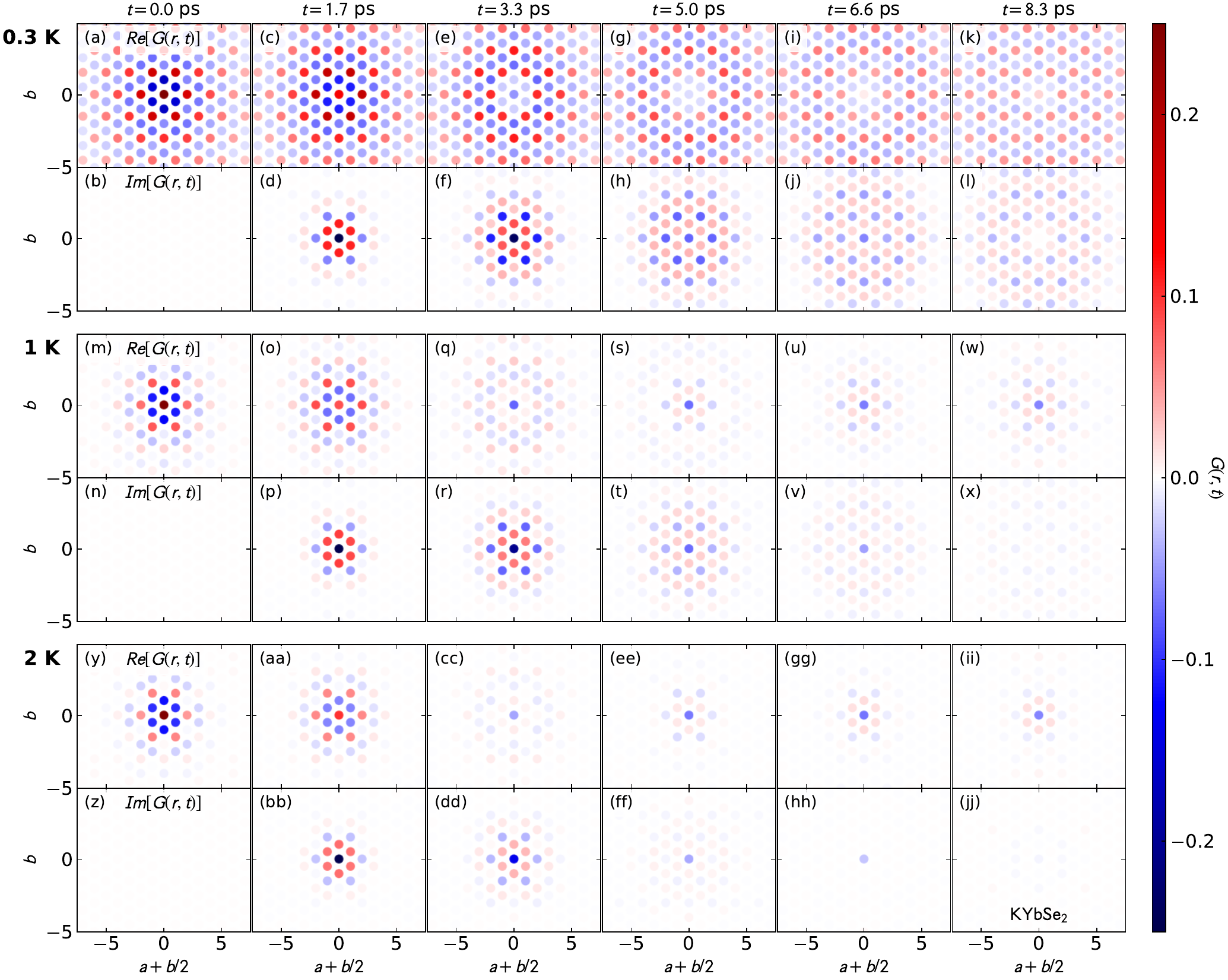} 
    \caption{Constant time slices of $G(r,t)$ in KYbSe$_2$. Panels (a)-(l) show the data at 0.3~K, panels (m)-(x) show the data at 1~K, and panels (y)-(jj) show the data at 2~K. The top row in each temperature shows the real part of $G(r,t)$, and the bottom row shows the imaginary part. 
    }
    \label{fig:ConstantTime}
\end{figure*}

Figure \ref{fig:ConstantTime} shows constant time $G(r,t)$ slices of KYbSe$_2$ for a different view of the time evolution. At 0.3~K the system shows a propagating wave moving outward from $(0,0)$, but the wavefront largely preserves the same spin correlation pattern. At 1~K and 2~K the spin correlation pattern inverts from the $t=0$ state, as noted in the main text. 

\section{Comparing theory and experiment}

The LSWT simulation performed at $T=0$ (second row of main text Fig. 2) looks dramatically different than  KYbSe$_2$ in $G(r,t)$, with the exception of reproducing the non-uniform ${Im}[G(r,t)]$ light cone onset. Otherwise, LSWT predicts a complex pattern of oscillations above the light cone which are not observed in experiment. This discrepancy is not surprising, as linear magnon dynamics have already been shown to fail in describing KYbSe$_2$ \cite{Scheie2024_KYS_SI}. 

The Schwinger Boson simulation performed at $T=0$ (third row of main text Fig. 2) comes the closest to reproducing the experimental $G(r,t)$, with a short-time $r=0$ inversion, and a close match to both real and imaginary components for $|r|>0$. This suggests that the assumptions behind the Schwinger Boson description correspond closely to  KYbSe$_2$'s actual behavior. 

The LL simulations with $k_B T = J_1/16$ (fourth row of main text Fig. 2), surprisingly, also give a close match to experiment. Here we also observe short-time $r=0$ inversion (but only in $Re[G(r,t)]$) and a close correspondence to the experimental ${Im}[G(r,t)]$ (though the correlations do not extend as long in time as the Schwinger Boson model or experiment). There is also close correspondence to the higher temperature data, shown in Fig. \ref{fig:LLD_Tdep}.

The RPA calculations performed at $T=0$, (fifth row of main text Fig. 2) show marginal agreement with experimental $G(r,t)$. Although the RPA $G(r,t)$ contains many rich and interesting features and the $S(q,\omega)$ shows diverging intensity at $K$, and the imaginary component does strongly resemble experiment, the real component does not resemble the KYbSe$_2$ spin correlations in detail. 

Finally, the MPS calculations performed at $T=0$ (sixth row of main text Fig. 2) show reasonable agreement with $S(q,\omega)$ but poor agreement $G(r,t)$ because of the finite-size gap induced by the small system size. This creates a ringing effect which modulates the spin correlations at even small distances and times. 

From this comparison it is obvious that Schwinger Boson calculations reproduce the KYbSe$_2$ spin correlations the best, followed closely by LL. Although this was reasonably clear from $S(q,\omega)$ alone, the $G(r,t)$ analysis confirms we can effectively rule out LSWT and $U(1)$ RPA as descriptions for KYbSe$_2$.

\section{Quantum correlation length}

\begin{figure}
    \includegraphics[width=0.49\textwidth]{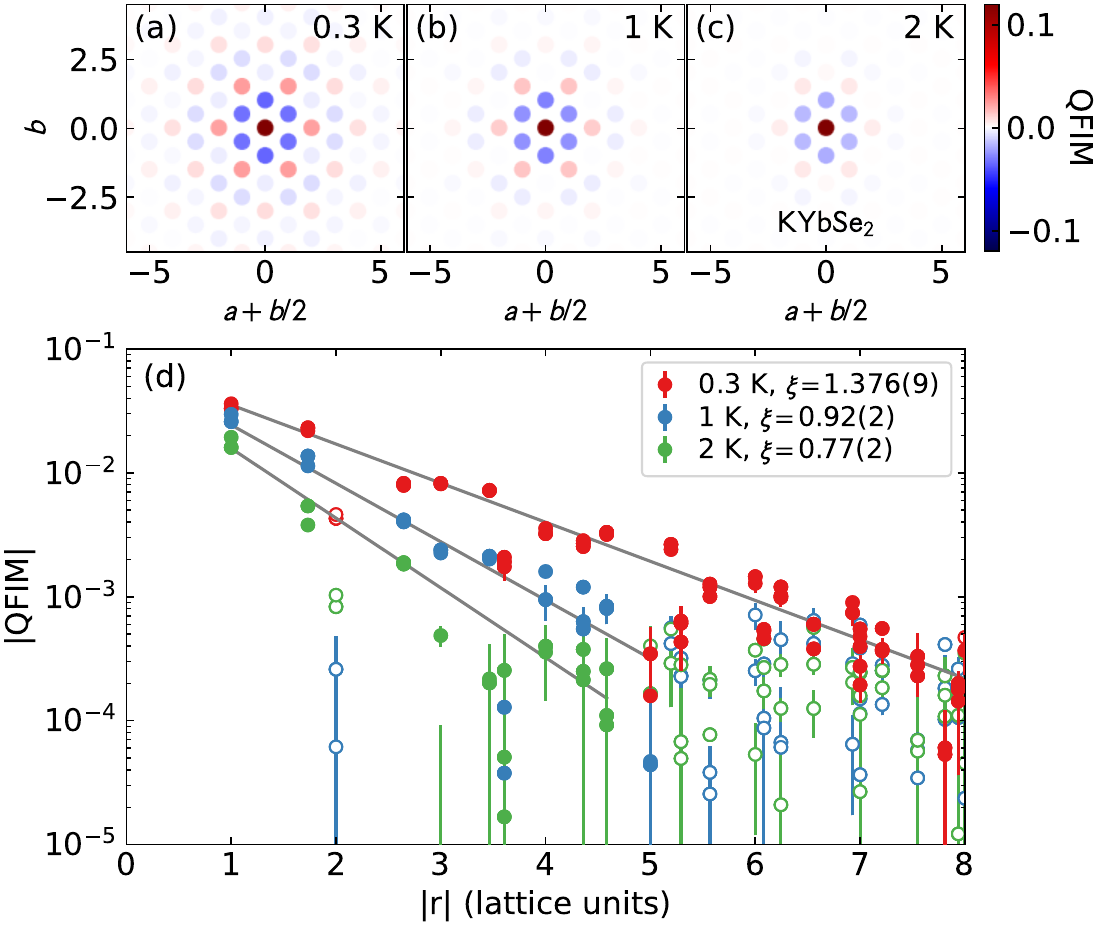} 
    \caption{Quantum Fisher Information Matrix (QFIM) of KYbSe$_2$. Panels (a)-(c) show colormaps of the QFIM at the measured temperatures. Note that the third neighbor correlations nearly vanish at 0.3~K and entirely vanish at 1~K. Panel (d) shows the absolute value of the correlations as a function of distance, and an exponential fit to extract a quantum correlation length (which excludes the sites where correlations are much smaller than neighbors). Closed symbols indicate fitted points, open symbols indicate excluded points. Error bars indicate one standard deviation uncertainty. 
    }
    \label{fig:QFIM}
\end{figure}

Having obtained the spin-correlations as a function of distance, it is possible to compute spatially-dependent Quantum Fisher Information Matrix (QFIM) \cite{scheie2024reconstructing_SI}. 
We perform a Fourier transform in space but carry out the quantum Fisher information integral in frequency 
\begin{equation}
    {\rm QFIM}[{\bf r}] =  \frac{1}{\pi} \int_0^\infty \mathrm{d}\left( \hbar \omega\right)\, \tanh \left( \frac{\hbar \omega}{2 k_B T} \right) \chi^{\prime\prime}_{\alpha \alpha} \left({\bf r}, \omega \right)~,
\label{eq:C}
\end{equation}
where $\chi^{\prime\prime} \left({\bf r}, \omega \right)$ is the dynamic susceptibility obtained from $S(q,\omega)$ via $ \chi_{\alpha \alpha}''\left( {\bf Q}, \omega \right)	=
 \pi \left( 1-e^{-\hbar\omega/k_B T} \right) S_{\alpha \alpha}({\bf Q},\omega)$ \cite{SCHEIE2025100020_SI} via a Fourier transform in reciprocal space: $\chi^{\prime\prime} \left({\bf r}, \omega \right) = \int d {\bf Q} e^{i \pi {\bf Q} \cdot {\bf r}} \chi^{\prime\prime} \left({\bf Q}, \omega \right) $. 
In Fig. \ref{fig:QFIM} we show the QFIM at the three measured temperatures. The upper panels show the spatial pattern, and the bottom panel shows the absolute value value QFIM versus absolute value of distance.

The QFIM reveals a non-uniform spatial decay of correlations. For instance, the third neighbor sites have very small QFIM at $T=0.3$~K (and essentially zero QFIM at $T=1$~K), whereas the fourth neighbors have a strong signal at both temperatures. This aperiodic QFIM pattern [which also is present in $G(r,t=0)$] persists to quite far distances. In the next section, we show that this pattern is associated with the lines of intensity connecting $K$ and $M$ in $S(q)$, and proximity to the classical degenerate point $J_2/J_1=1/8$. 
 
Taking the absolute value of QFIM versus distance, we observe a temperature-dependent quantum correlation length in KYbSe$_2$. The quantum correlations decay with an exponential envelope, from which we can define and fit an overall quantum correlation length as shown in Fig. \ref{fig:QFIM}. Like the case of the 1D spin chain \cite{scheie2024reconstructing_SI}, this quantum correlation length is much shorter than the total correlation length, and strongly varies with temperature.

\section{Fourier analysis}

\subsection{Static correlation pattern}

\begin{figure}
    \includegraphics[width=0.48\textwidth]{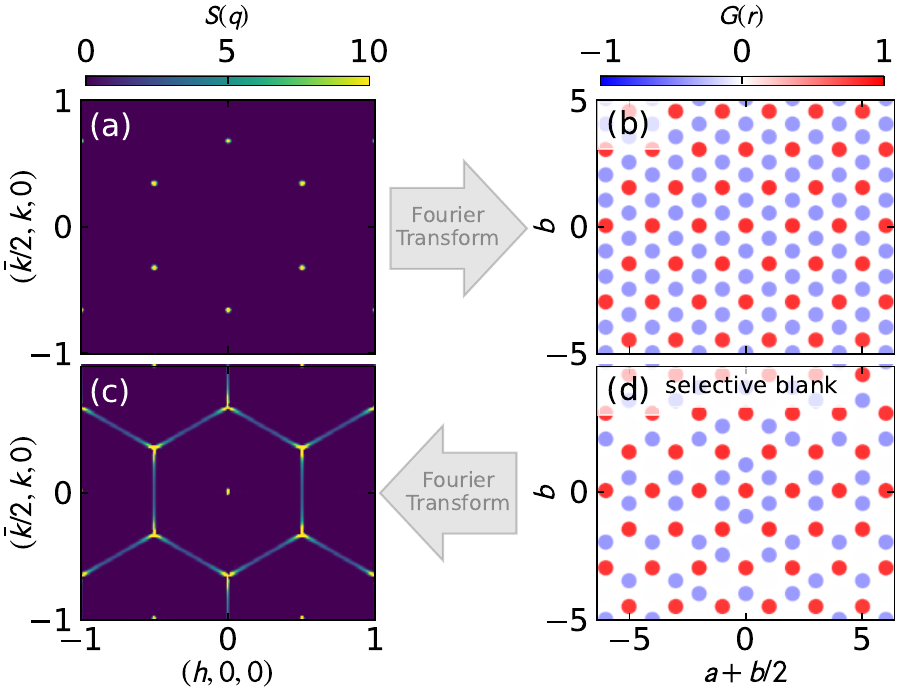} 
    \caption{Fourier analysis of 2D triangular spectrum. (a) shows the $S(q)$ of a long-range $120^{\circ}$ order, with the $G(r)$ signal in panel (b) obtained by Fourier transform. Panel (d) shows the same $G(r)$ with the sites selectively set to zero to match the pattern observed in KYbSe$_2$. Panel (c) shows the spectrum Fourier transformed back into reciprocal space, revealing lines of intensity connecting the $K$ and $M$ points.}
    \label{fig:FourierAnalysis}
\end{figure}

In the Quantum Fisher Information Matrix (QFIM) calculation in Fig. \ref{fig:QFIM}, as well as the constant-time plots in Fig. \ref{fig:ConstantTime}, there is an aperiodic pattern of suppressed correlations at the third neighbor sites, sixth neighbor sites, etc. To understand the origin of this pattern, we performed Fourier analysis on simulated spin correlations, as shown in Fig. \ref{fig:FourierAnalysis}. Beginning with a simple $120^{\circ}$ order, we transform $S(q)$ into $G(r)$ (ignoring the time dimension), and selectively set the spin correlations to zero along the pattern observed in main text Fig. 5. We then took an inverse Fourier transform back to reciprocal space.

The result shows a pattern of sharp Bragg peaks at $K=(1/3,1/3)$ transformed into a pattern of hexagonal streaks of intensity after selectively suppressing the spin correlations.  These lines of intensity connect $K$ (associated with $120^{\circ}$ order) and $M$ (associated with stripe order), the ordering wavevectors which compete and which are separated by the quantum spin liquid phase. Indeed, matrix product calculations have shown that the quantum spin liquid phase is characterized by lines of intensity connecting the $K$ and $M$ points \cite{Sherman_2023_Spectral_SI}. 

Thus the peculiar pattern of suppressed spin correlation observed in KYbSe$_2$ is associated with the lines of intensity connecting $K$ and $M$. However, it must be pointed out that there is nothing especially quantum-mechanical about this intensity pattern, as the lines of intensity can be reproduced at the classical degenerate point $J_2/J_1 = 1/8$ where $120^{\circ}$ order can be transformed to stripe order with zero energy cost. Be that as it may, this intensity pattern is definitely evidence that the system is close to the $T=0$ phase boundary, and thus KYbSe$_2$ is close to the spin liquid on the quantum phase diagram. 

\subsection{Nonlinear light cones}

\begin{figure*}
    \includegraphics[width=0.9\textwidth]{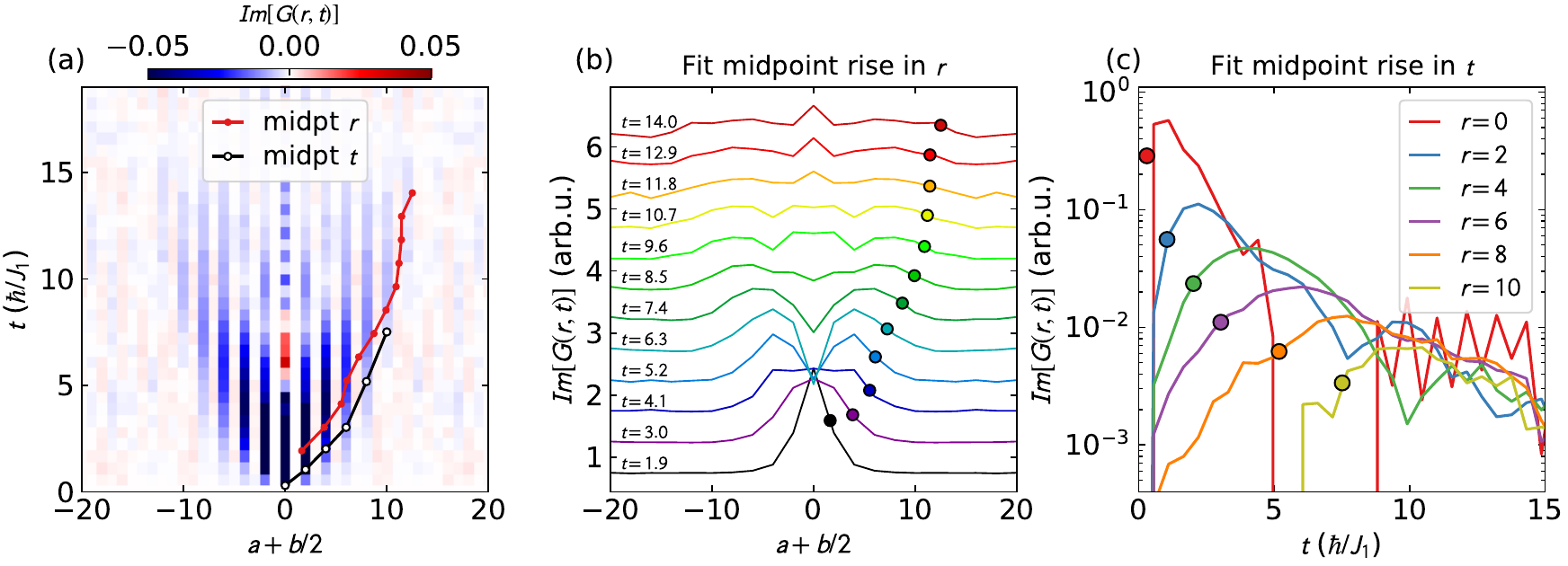} 
    \caption{Methods for fitting the light cone from KYbSe$_2$ data. Panel (a) shows the imaginary $G(r,t)$ along the second neighbor direction. With a reduced colormap range, the nonlinear light cone is directly apparent in the data. Panel (b) shows the fitted midpoints in the bound of the light cone in $r=(a+b/2)$ (indicated by circles). Panel (c) shows the midpoints of the rise in time (also indicated by circles). The fitted midpoints are plotted overtop panel (a). Although these estimates of the light cone boundary do not agree, they both show a distinct nonlinearity bending upwards. 
    }
    \label{fig:altfits}
\end{figure*}

In the main text we show that KYbSe$_2$ has a nonlinear light cone which none of the theoretical methods successfully reproduce. In Fig. \ref{fig:altfits} we show details of the method used to fit the light cone by the midpoint in the rise in time, as well as an alternate definition of the light-cone defined by fitting versus space. Both show a distinct nonlinearity, but for simplicity we focus in the main text on the nonlinearity found by fitting versus time. 

\subsubsection{Experimental broadening}

Here we explore whether experimental broadening can produce a nonlinear light cone as observed in KYbSe$_2$. 
In reality, the neutron data includes experimental broadening which will affect the $G(r,t)$ signal. 
Furthermore, the neutron scattering data used to generate $G(r,t)$ in this study comes from two incident energies stitched together: below  $\hbar \omega  = 0.2$~meV the data is from $E_i = 1.55$~meV, and above $\hbar \omega  = 0.2$~meV the data is from $E_i = 3.32$~meV \cite{Scheie2024_KYS_SI} (the use of lower $E_i$ gives a higher resolution close to the elastic line \cite{stone2014comparison_SI}).

\begin{figure*}
    \includegraphics[width=0.9\textwidth]{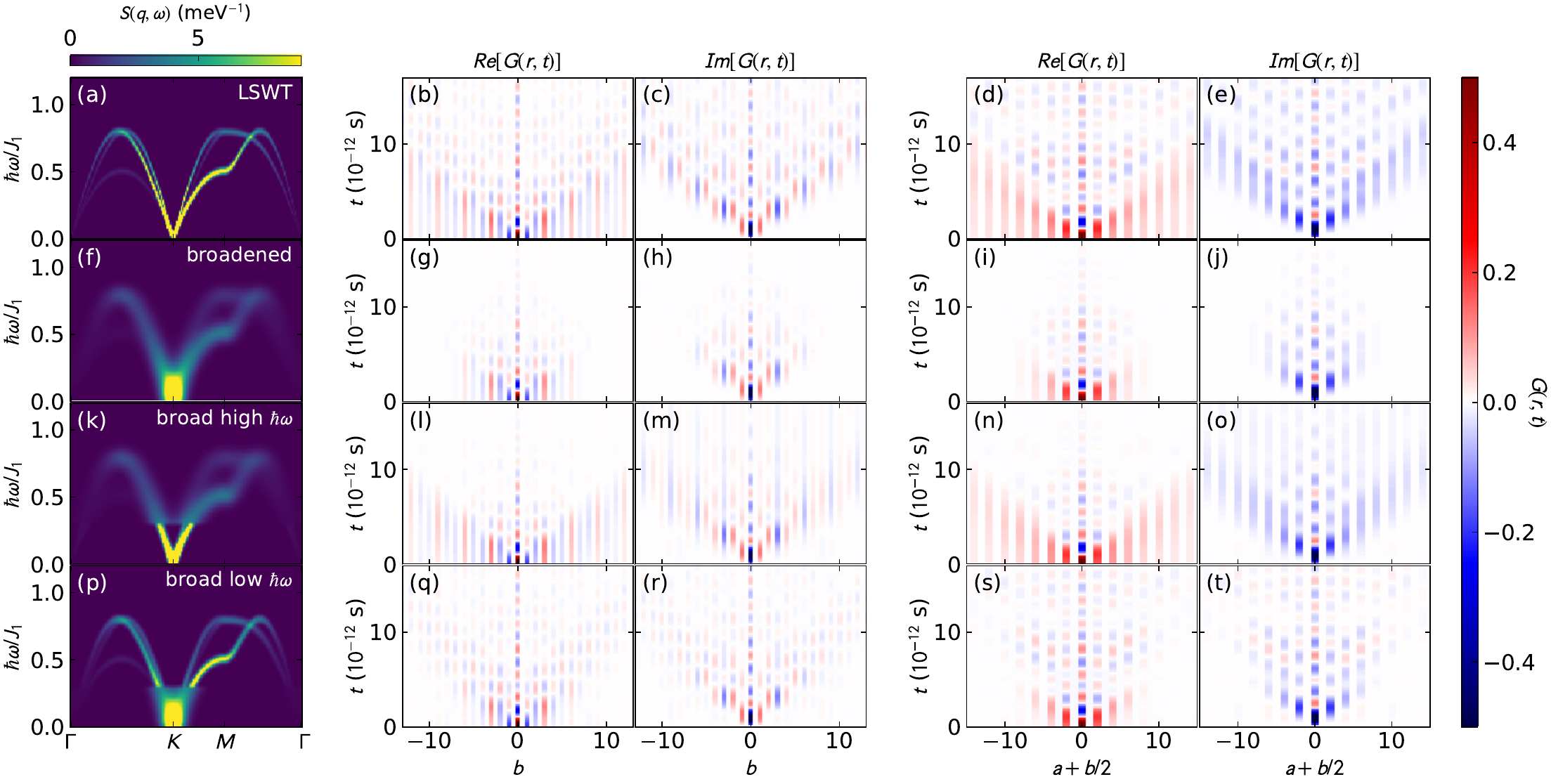} 
    \caption{Linear spin wave theory (LSWT) $G(r,t)$ calculated from broadened spectra. The top row (a)-(e) shows the $S(q,\omega)$ and $G(r,t)$ from the original sharp LSWT spectrum. The second row (f)-(j) shows the results from uniform Gaussian broadening, which suppresses the $G(r,t)$ signal to short times and distances but does not generate a nonlinear light cone. The third row (k)-(o) shows the results from the high energy spectrum being broadened but the low energy spectrum remaining sharp. This suppresses the spin oscillations above the light cone, but does not produce an upward bend in the light cone itself. The fourth row (p)-(t) shows the inverse from above: low energy broad spectrum but a high energy sharp spectrum (experimentally unrealistic for KYbSe$_2$). In this case there is a nonlinearity induced, but this neither resembles the real spectrum nor is experimentally realistic. This indicates the  KYbSe$_2$ nonlinear light cone is not an artifact of broadening. 
    }
    \label{fig:LSWT_broadening}
\end{figure*}

To examine whether these experimental effects might cause a linear light cone to become nonlinear, we applied Gaussian broadening to the linear spin wave theory (LSWT) calculated spectrum shown in Fig. \ref{fig:LSWT_broadening}. Applying uniform broadening merely suppresses the $G(r,t)$ at large distances and times (as expected for a convolved Fourier transform). Meanwhile, if we apply an energy-dependent broadening such that the low-energy modes are sharper than the high energy modes (as in our experimental data), there is no nonlinearity produced, but a marked loss of clear spin oscillations at long times. However, if we apply broadening to the low energy modes only, we find that the light cone does become nonlinear because the short time spectra are suppressed less than the long-time spectra. That said, given that (i) this is the inverse of our experimental situation and (ii) the measured KYbSe$_2$ spectrum is definitely more diffuse at high energies, we do not consider this a valid explanation for the nonlinear KYbSe$_2$  light cone. 
From this exercise we conclude that the observed nonlinear light cone is not from experimental broadening effects, and is an intrinsic feature of KYbSe$_2$.

\subsubsection{Nonlinear LSWT light cones}

\begin{figure*}
    \includegraphics[width=\textwidth]{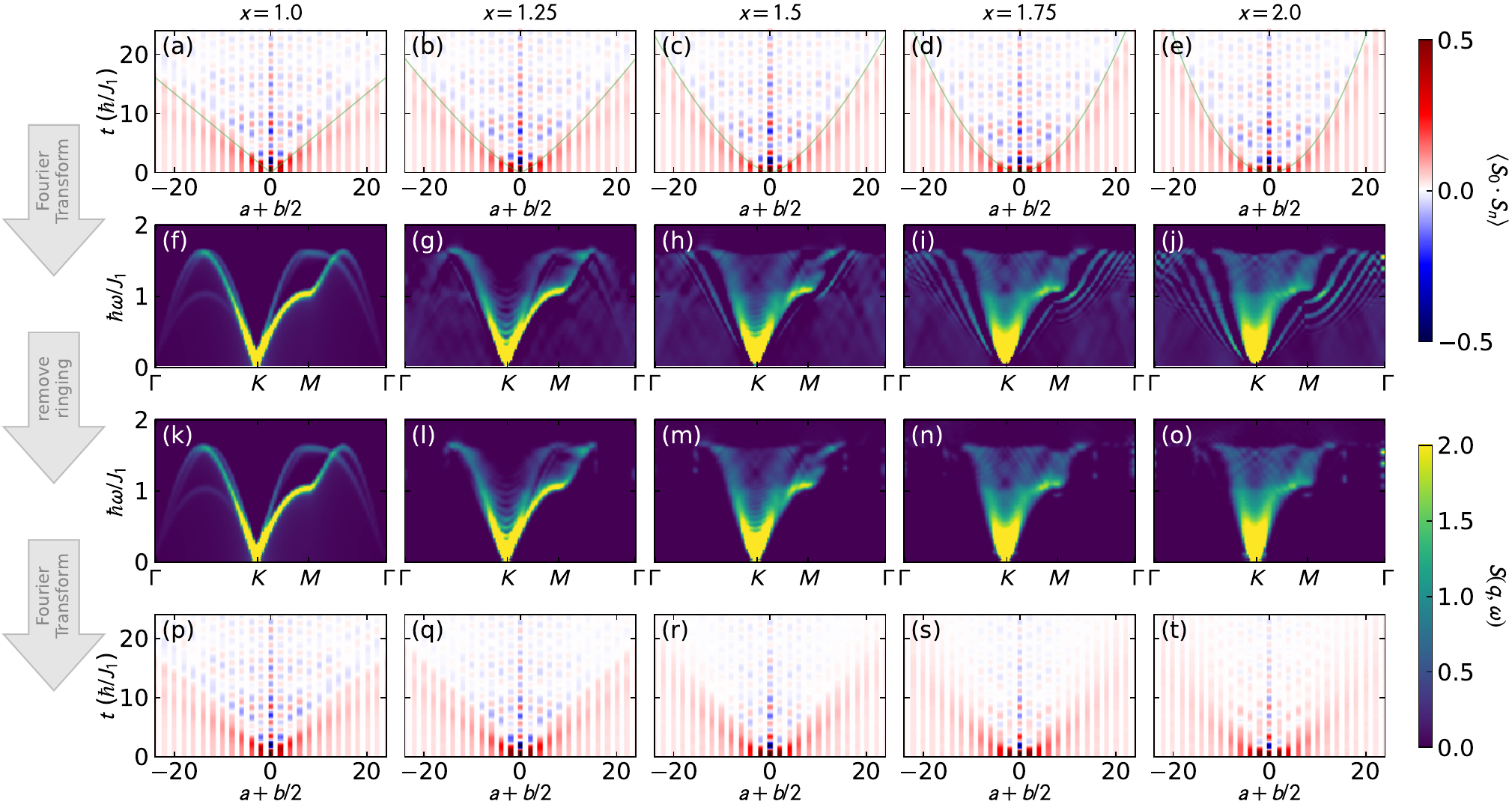} 
    \caption{LSWT $G(r,t)$ manually distorted so that the light cone onset matches dynamic exponents $1 \leq x \leq 2$ (a)-(e) where the green line shows the light cone, and the resulting $S(q,\omega)$ from Fourier transforming back to reciprocal space and frequency (f)-(j). Sharp excitations for $x=0$ evolve to a diffuse continuum in the ballistic limit.     
    The bottom two rows show the same $S(q,\omega)$ with the ringing artifacts manually removed (k)-(o), and transformed back to $G(r,t)$ (p)-(t). Removing the ringing features removes the sharp bound to the light cone, but  the nonlinear onset persists. 
    }
    \label{fig:LightCone_LSWT_ringing}
\end{figure*}

In principle, nonlinear light cones in $G(r,t)$ have some signature in $S(q,\omega)$ that should be identifiable. 
To examine the effect of a nonlinear light cone on $S(q,\omega)$, we take the LSWT calculation for $120^{\circ}$ order, Fourier transform to $G(r,t)$, and distort the light cone to match various dynamic exponents $x$ between $x=1$ (ballistic) and $x=2$ (diffusive). This is done by shifting the $G(r,t)$ in time for each $r$, and extending the $t=0$ value to finite $t$ to fill any gaps. We then transform back to $S(q,\omega)$ and plot the intensity along high-symmetry directions. The results are shown in Fig. \ref{fig:LightCone_LSWT_ringing}(a)-(j). 
This exercise is not general because the LSWT solution has a unique pattern of spin correlation above the light cone, but it still gives a rough understanding of how a spectrum evolves from ballistic to diffusive.


The two features of note in Fig. \ref{fig:LightCone_LSWT_ringing}(f)-(j) as $x \rightarrow 2$ are (1) that the spectrum evolves from  sharp features to a broad continuum, and (2) the lower bound to the continuum evolves from linear to nonlinear. 
There is also a ringing feature in the Fourier transform (seen by the ripples outside the continuum). We explore this in Fig. \ref{fig:LightCone_LSWT_ringing}(k)-(t) by manually removing the spectral weight outside the continuum and transforming back to $G(r,t)$. We find that removing the ringing blurs the bound of the light cone (this makes sense as sharp features generically produce oscillatory behavior in a Fourier transform), but the generic nonlinearity persists. We therefore  associate a nonlinear $G(r,t)$ light cone with a continuum in $S(q,\omega)$ that has a nonlinear bound at low energies. 


\section{LL dynamics}

\begin{figure*}
    \includegraphics[width=\textwidth]{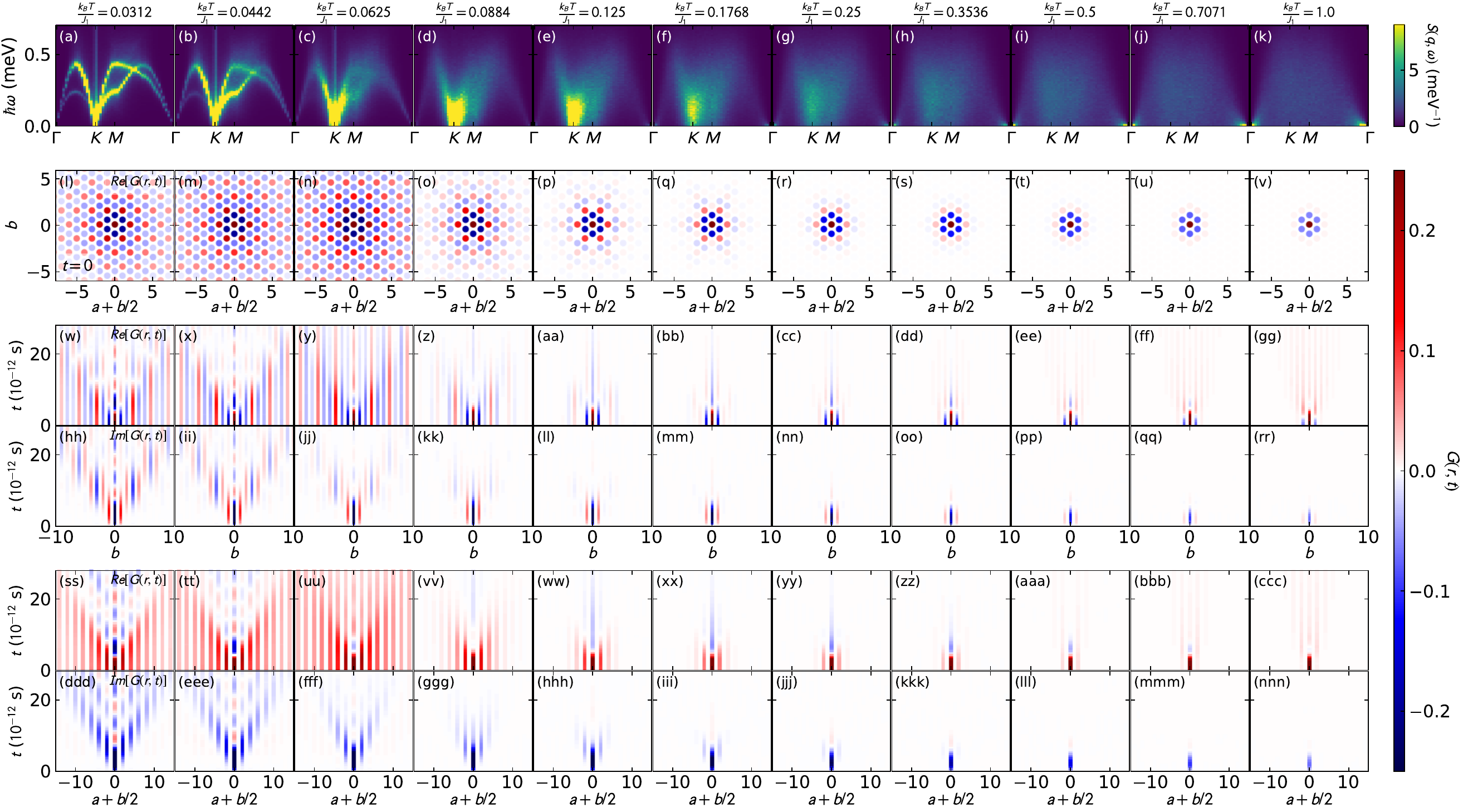} 
    \caption{Van Hove correlations for Landau-Lifshitz (LL) dynamics at various temperatures with $J_2/J_1 = 0.044$ (the fitted value for KYbSe$_2$ \cite{Scheie2024_KYSNYS_SI}). Near the ordering temperature $k_B T = 0.0625 J_1$, the system displays the long-time correlations, lack of $r>0$ oscillations, and weak light cone observed in KYbSe$_2$ at low $T$. 
    Figure \ref{fig:LLD_Tdep_b} shows this with finer temperature resolution around $T_N$. 
    }
    \label{fig:LLD_Tdep}
\end{figure*}

\begin{figure*}
    \includegraphics[width=0.7\textwidth]{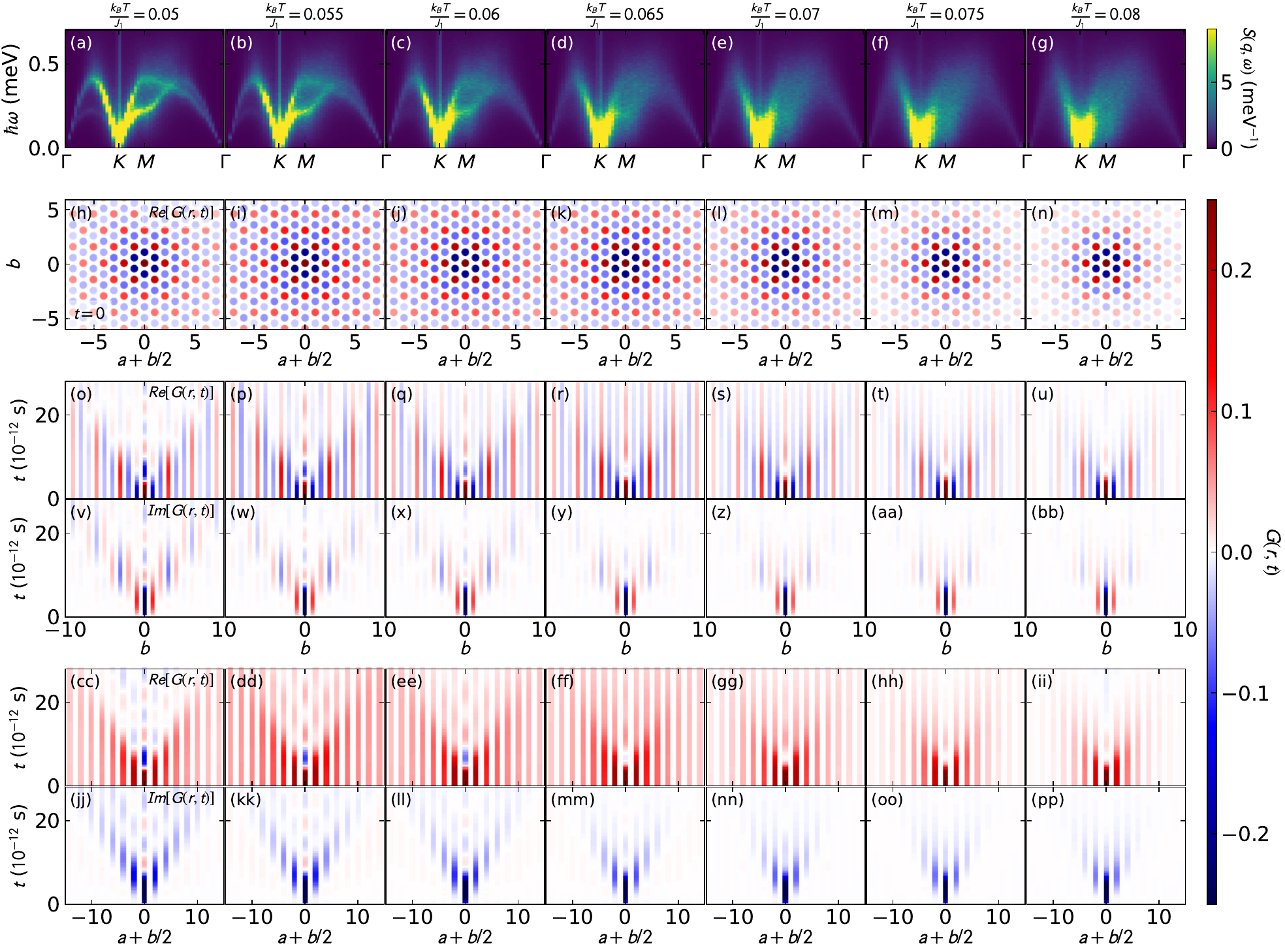} 
    \caption{Van Hove correlations for Landau-Lifshitz (LL) dynamics with $J_2/J_1 = 0.044$ for temperatures around $T_N$, plotted similar to Fig. \ref{fig:LLD_Tdep}. 
    }
    \label{fig:LLD_Tdep_b}
\end{figure*}

\begin{figure*}
    \includegraphics[width=\textwidth]{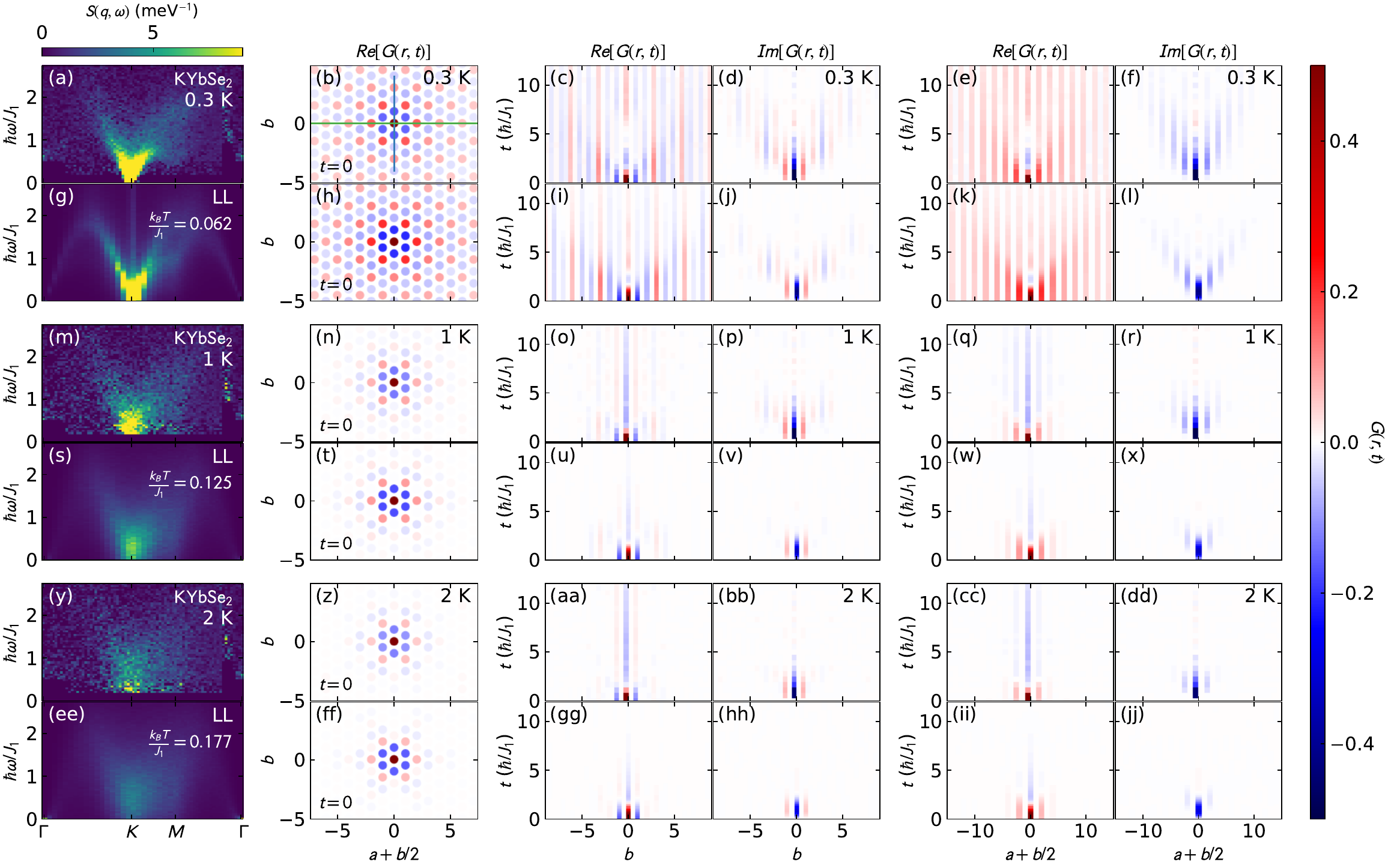} 
    \caption{Van Hove correlations for KYbSe$_2$ and Landau-Lifshitz (LL) dynamics at various temperatures. The arrangement of panels is the same as main text Fig. 2. 
    Although the match between experiment and LL is imperfect, the correspondence is remarkably close at all temperatures, especially at 1~K and 2~K. 
    }
    \label{fig:KYS-LLD}
\end{figure*}

The triangular lattice Landau Lifshitz (LL) dynamics  calculations were performed on a system of 9,216 spins ($48 \times 48 \times 2$ unit cells) with periodic boundary conditions. 
Temperature evolution of the triangular lattice $S(q,\omega)$ and $G(r,t)$ is shown in Fig. \ref{fig:LLD_Tdep}, focused in at temperatures around $T_N$ in Fig. \ref{fig:LLD_Tdep_b}, and compared with KYbSe$_2$ experiment in Fig. \ref{fig:KYS-LLD}. As noted in the main text, the agreement at finite temperatures between theory and experiment is remarkable. 

Because the system is finite, long-range order occurs even though in the infinite lattice limit order only happens at $T=0$. Nevertheless, if we take this finite-size artifact as an approximation for the 3D and anisotropic exchange that produces long range order in the real  KYbSe$_2$ system, we can observe the thermal evolution of $G(r,t)$ with much finer steps than is practical in experiment. As noted in the main text, the long-time persistent $Re[G(r,t)]$ appears at the transition temperature, indicating a divergence in the timescale of fluctuations. 

\subsection{Emulating quantum phase diagrams}

One of the curious features of the LL simulations is how well they mimic the behavior of KYbSe$_2$ at all temperatures.  
This effect has been seen before. Studies of the Kitaev model showed that classical dynamics with a semiclassical intensity correction  \cite{Huberman_2008_SI} reproduces the full quantum solution with remarkable accuracy---except at low frequencies where long-range quantum dynamics are most manifest \cite{Samarakoon_2017_SI,Samarakoon_2018_SI,Franke_2022_SI,Kim_2025_Emulation_SI}, which here is a crossover in time: short-time dynamics are mimicked by classical simulations, but not longer times.

\begin{figure*}
    \includegraphics[width=\textwidth]{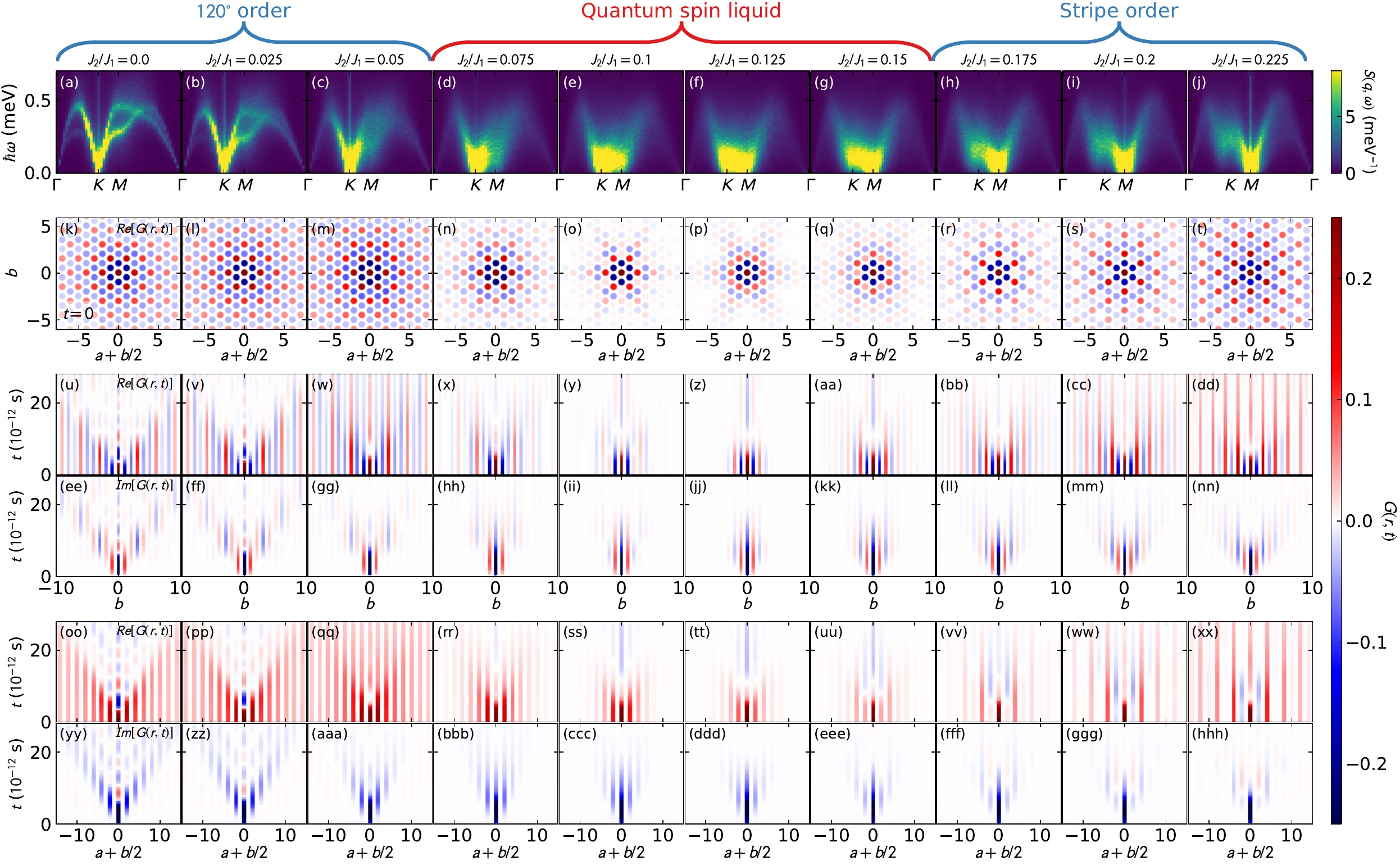} 
    \caption{Emulating quantum effects with thermal fluctuations: Van Hove correlations for Landau-Lifshitz (LL) dynamics at $k_B T = J_1/16$ at various values of $J_1/J_1$. In between the regions of stripe AFM order and $120^{\circ}$ order is a region of magnetic disorder, which mimics the quantum spin liquid phase diagram to a remarkable degree of accuracy. 
    }
    \label{fig:KYS-LLD2}
\end{figure*}

Interestingly, we can emulate the quantum behavior of the 2D triangular lattice by tuning the LL simulations through the entire $J_2/J_1$ phase diagram, shown in Fig. \ref{fig:KYS-LLD2}. In this case, a small finite temperature $k_B T = J_1/16$ reproduces the triangular $J_2/J_1$  quantum phase diagram \cite{PhysRevB.92.041105_SI,PhysRevB.92.140403_SI,PhysRevB.93.144411_SI,PhysRevB.94.121111_SI,PhysRevB.95.035141_SI,PhysRevB.96.075116_SI,PhysRevLett.123.207203_SI} with impressive accuracy, tuning from $120^{\circ}$ to stripe order with an intermediate non-magnetically-ordered liquid regime at $0.06 \lesssim J_2/J_1 \lesssim 0.16$. 
The $J_2/J_1 = 0.05$ $G(r,t)$ simulations closely resemble KYbSe$_2$ $G(r,t)$, while moving further from this point in either direction produces a qualitatively different signal. 
This confirms that (i)  LL is an effective quantum emulator able to reproduce quantum phase diagrams \cite{Samarakoon_2022_SI}, and (ii) KYbSe$_2$ lies extremely close to the boundary with a quantum liquid phase. 

Something interesting in Fig. \ref{fig:KYS-LLD2} is that the long-time persistent $Re[G(r,t)]$ signal as observed in KYbSe$_2$ is only found on the boundary between magnetic order and quantum spin liquid. Deep in the ordered phase or deep within the quantum spin liquid, the light cone exhibits dynamic, flipping spins (red-blue oscillations in the $G(r,t)$ plots). Near the phase boundary, the $t=0$ pattern persists to very long times---even above the light cone. 
As noted above, this occurs also at the phase boundary in temperature between magnetic order and paramagnetism. 
Thus this long-time persistent $G(r,t)$ is a signature of critical behavior, wherein the relaxation timescale diverges.  Intuitively, this is not surprising, as critical fluctuations will appear as frequency-scale-free signatures in $S(q,\omega)$ which transform to time-scale-free decays in $G(r,t)$. 
This behavior in KYbSe$_2$ at low temperature confirms that KYbSe$_2$ is near a quantum critical phase boundary separating $120^{\circ}$ and a quantum spin liquid. 

Another interesting feature revealed by the emulated phase diagram Fig. \ref{fig:KYS-LLD2} is that within the spin liquid regime, the $Re[G(r,t)]$ exhibits spin-pattern-inversion above the light cone. This was also observed at 1~K and 2~K in KYbSe$_2$. This indicates that KYbSe$_2$ at finite temperature is indeed acting as the quantum spin liquid. Unlike the case of the 1D spin chain where spinons are heuristically understood as propagating domain walls \cite{Scheie2022_SI}, a more detailed understanding of the 2D triangular spinons is not forthcoming from $G(r,t)$. Nevertheless, these data show that the propagation of spinons (at least at short timescales) produces an inverted spin correlation pattern. 

Thus, despite the fact it is a classical approximation, finite-$T$ LL simulations provide useful insight to interpreting $G(r,t)$ from KYbSe$_2$.  The fact that this works so well is a consequence of the quantum-to-classical crossover  \cite{Huberman_2008_SI,Samarakoon_2017_SI,Samarakoon_2018_SI,Franke_2022_SI,Dahlbom_2024_Quantum_SI}. Here we confirm that this crossover also occurs in \textit{time}: at short times, classical simulations can reproduce the behavior of quantum materials, but they fail at long times where the relevant energy scales are small. 
Thus, at temperatures slightly above $T_N$ (where our experiments are performed), KYbSe$_2$ is expected to lie in a renormalized classical regime, in which thermal fluctuations strongly affect long-time dynamics while short-time correlations retain signatures of proximate quantum phases.

\subsection{Square lattice simulations} 

\begin{figure*}
    \includegraphics[width=\textwidth]{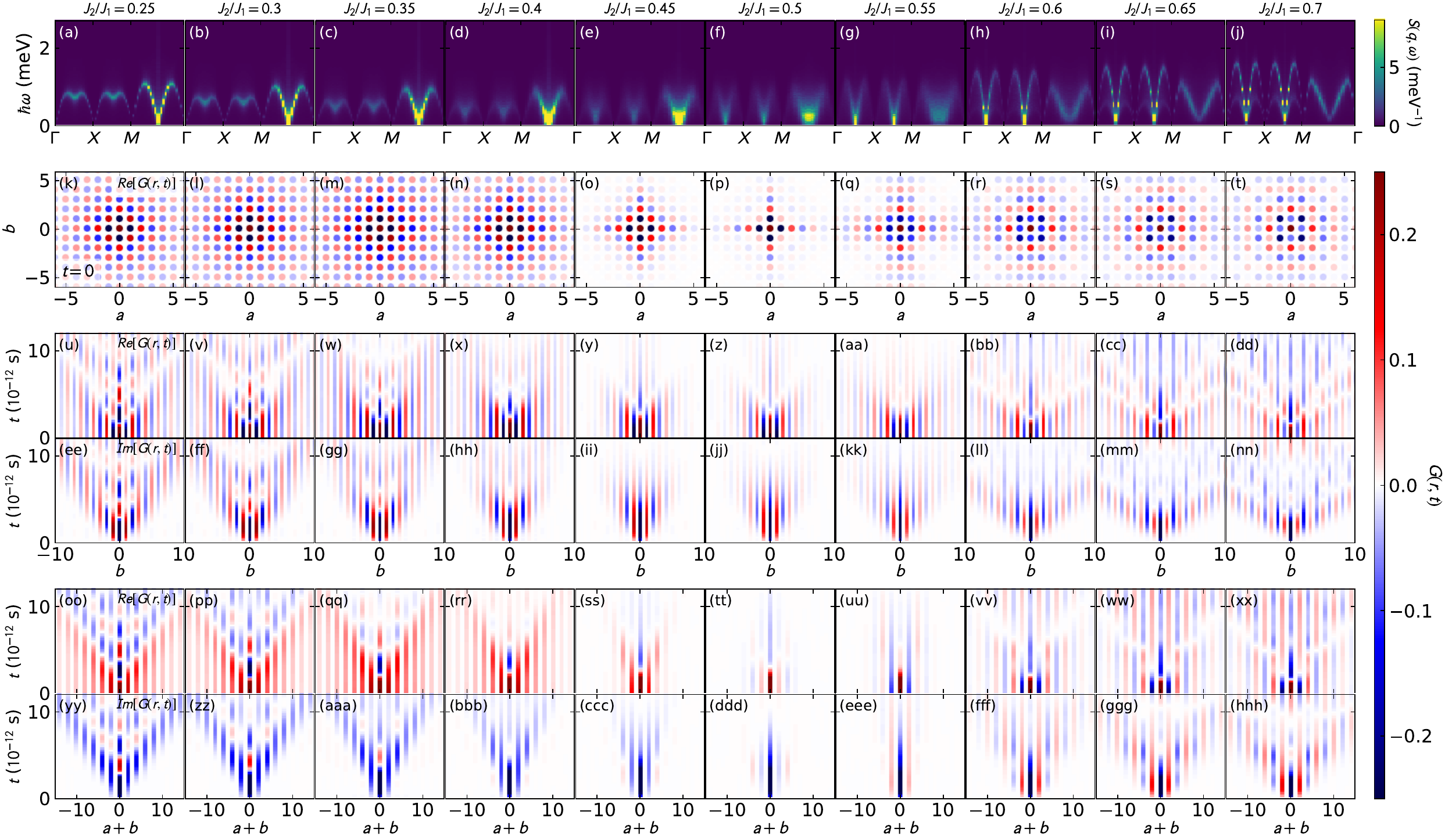} 
    \caption{Van Hove correlations for Landau-Lifshitz (LL) dynamics on a 3D square lattice with $k_B T = J_1/16$ for various values of second neighbor exchange $J_2/J_1$. Near  $J_2/J_1 = 0.5$ the long range order vanishes and the system exhibits only short ranged correlations. 
    }
    \label{fig:LLD_SquareVsJ2}
\end{figure*}

\begin{figure*}
    \includegraphics[width=\textwidth]{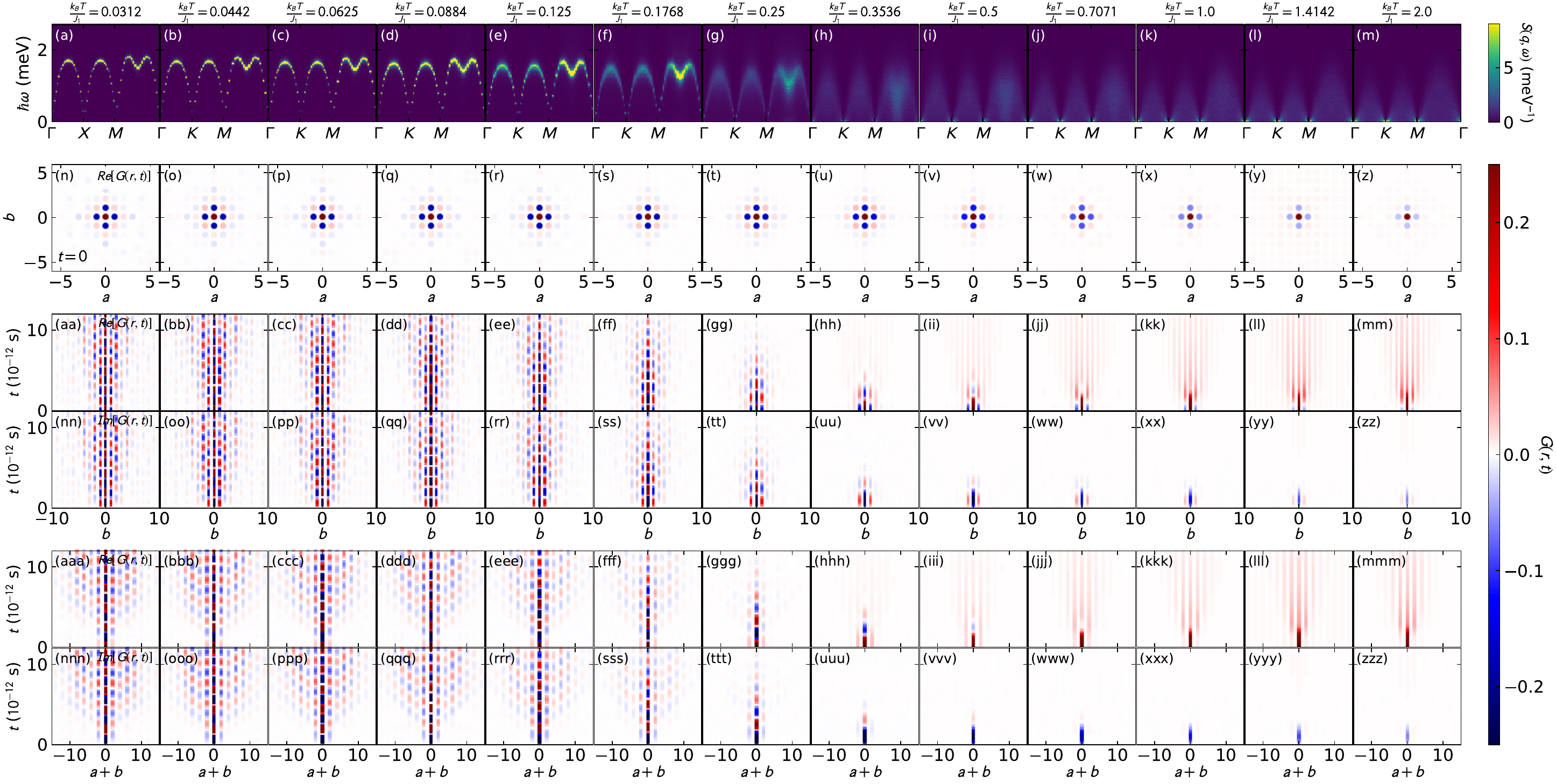} 
    \caption{Van Hove correlations for Landau-Lifshitz (LL) dynamics on a 3D square lattice with $J_2/J_1 = 0.044$ and $J_3/J_1 = 0.5$ as a function of temperature. Near the critical temperature, the ``quantum wake'' almost disappears and the $t=0$ pattern persists to longer times.  
    }
    \label{fig:LLD_SquareVsT}
\end{figure*}

These effects are not limited to the quasi-2D. In Figs. \ref{fig:LLD_SquareVsJ2} and \ref{fig:LLD_SquareVsT} we repeat the analysis for a simulated 3D square lattice ($48 \times 48 \times 8$ spins), where the nearest neighbor exchange $J_1$ is fixed, the out-of-plane exchange $J_3 = J_1/2$, and the $J_2$ exchange (diagonal across the squares in the plane) is varied. 

Similar to the 2D triangular case, there are competing phases on the square lattice as $J_2$ grows. In this case the competition is  between a simple $Q=(1/2,1/2)$ N\'eel state and a $Q=(0,1/2)$ stripe  order phase, separated by a degenerate point $J_2=J_1/2$. 
Figure \ref{fig:LLD_SquareVsJ2} shows the evolution of spin correlations as $J_2$ is tuned through this degenerate point, revealing a non-trivial evolution of the $G(r,t)$ signal. 

Figure \ref{fig:LLD_SquareVsT} shows the temperature evolution of the 3D square lattice with  $J_2$ set to a small value, such that the system is essentially unfrustrated. Near the onset of long range order and coherent magnons, the oscillations are severely damped and the long-time persistent spin correlations are observed with a single spin flip at $r=0$, very similar to the 2D triangular case in Fig. \ref{fig:LLD_Tdep}. This shows that the critical dynamics have universal features of $G(r,t)$, regardless of the dimensionality and level of frustration. 

\subsection{Simulating disorder}

\begin{figure*}
    \includegraphics[width=\textwidth]{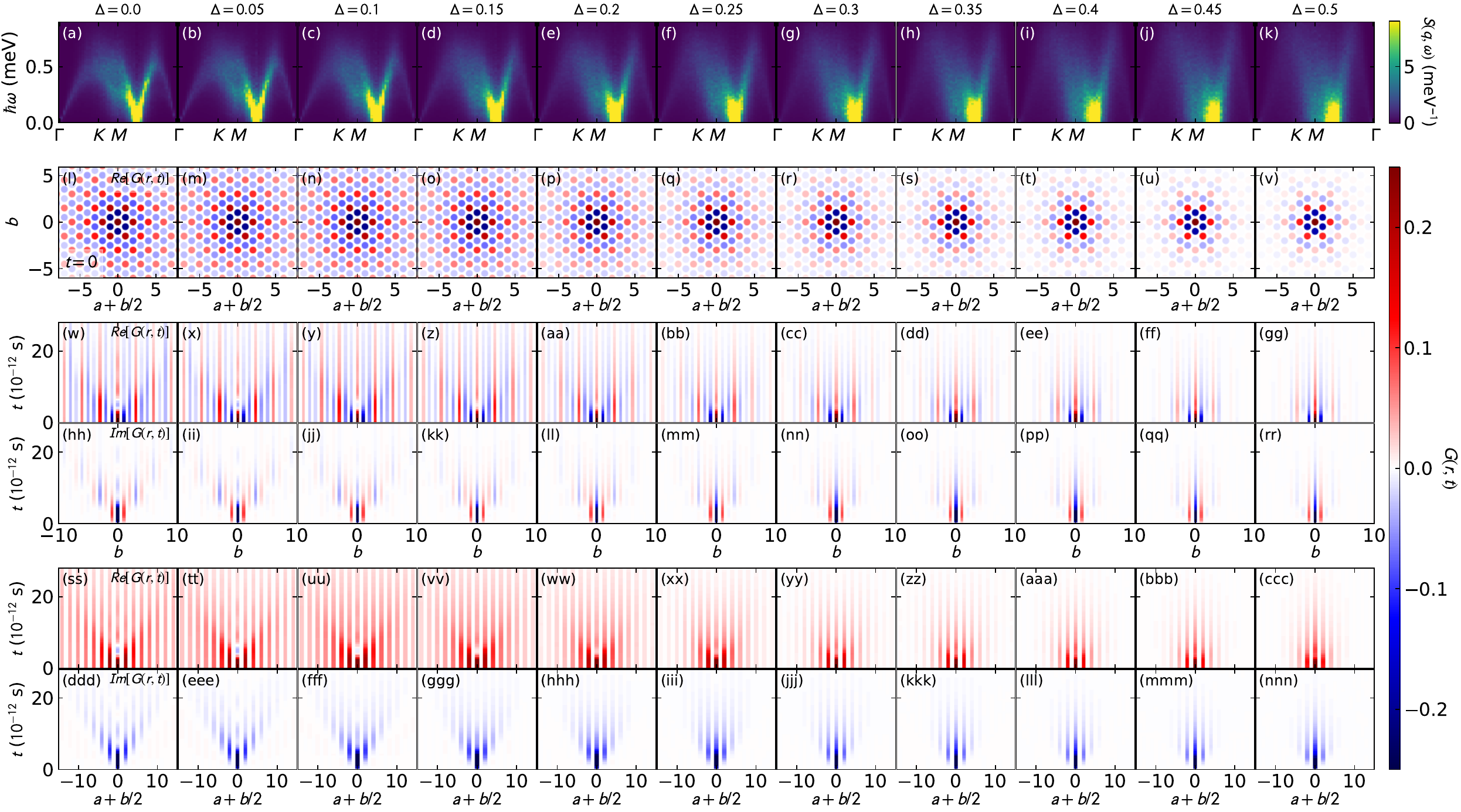} 
    \caption{Van Hove correlations for Landau-Lifshitz (LL) dynamics on the 2D triangular lattice with $J_2/J_1 = 0.044$ and $k_B T/ J_1 = 0.0625$ as a function of gaussian exchange disorder $\Delta$ (a relative quantity applied to both $J_1$ and $J_2$).  
    }
    \label{fig:LLD_Disorder}
\end{figure*}

It may be asked whether defects and disorder in the spin exchange Hamiltonian might cause a nonlinearity in the  $G(r,t)$ light cone. 
The full quantum treatment for such a problem is extremely difficult, but we use LL to approximate the effects of disorder. We simulated a $48 \times 48 \times 10$ lattice of $\approx 4.6 \times 10^4$ spins with random gaussian exchange disorder introduced to the model (as we neglect interplane coupling, this is effectively summing over 10 disorder configurations). 
Figure \ref{fig:LLD_Disorder} shows the Van Hove correlation functions for the 2D triangular lattice model with $J_2/J_1 = 0.044$ and $k_B T/ J_1 = 0.0625$ with distributions of spin exchange disorder. As disorder increases, the spectrum becomes more diffuse, and the light cones become restricted to shorter distances. Interestingly, the effects of disorder mimic finite temperature in the LL model, compare to Fig. \ref{fig:LLD_Tdep}. 

\begin{figure*}
    \includegraphics[width=0.98\textwidth]{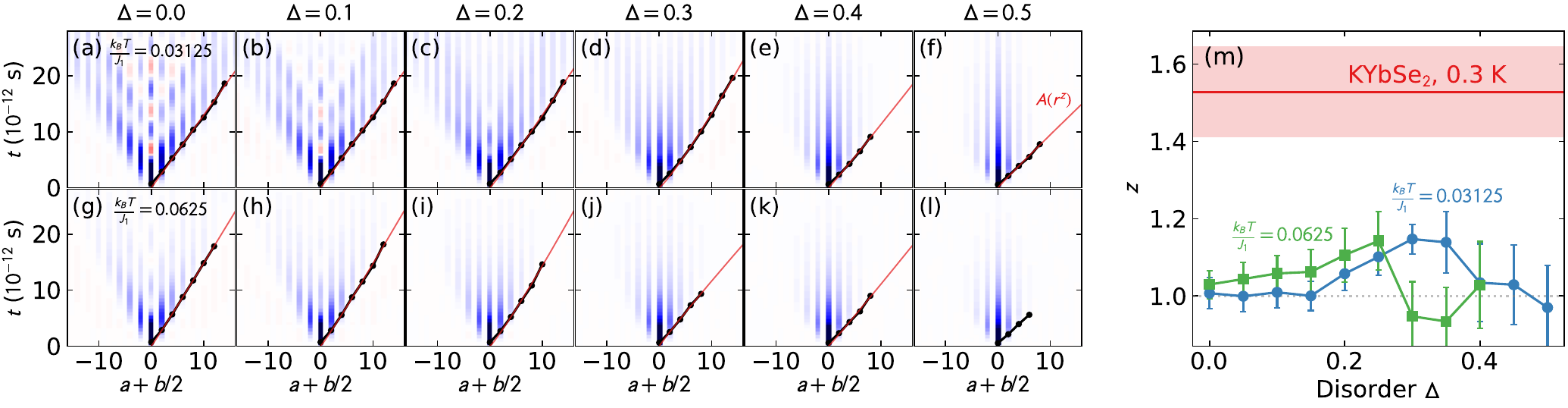} 
    \caption{Light cone onset along the second neighbor direction from LL simulations as a function of Gaussian exchange disorder. Panels (a)-(f) show the light cones at   $k_B T/ J_1 = 0.03125$, while panels (g)-(l) show the light cones at  $k_B T/ J_1 = 0.0625$. Black dots indicate the mid-point of the rise at different $r$, and the red line indicates a power law fit $A(r^z)$. 
    Panel (m) shows the fitted power law exponent for the different disorders. While disorder causes some slight nonlinearity to occur [c.f. panel (d)]. This is nowhere close to the fitted power law to KYbSe$_2$ at 0.3~K along the second neighbor direction $z=1.52(12)$. 
    Interestingly, the LL simulations show a monotonic increase in $z$ versus $\Delta$ up until a point where $z$ drops to 0 to within uncertainty. 
    }
    \label{fig:LLD_Disorder_LightCone}
\end{figure*}

Figure \ref{fig:LLD_Disorder_LightCone} shows the light cone analysis for the disordered LL simulations. Disorder appears to increase the light cone velocity (indicated by a shallower light cone onset and a steeper dispersion in $S(q,\omega)$), but it only introduces a mild nonlinearity (up to $z\approx1.2$) to the light cone which is far less dramatic than the nonlinearity observed in KYbSe$_2$ experiment.

\begin{figure*}
    \includegraphics[width=\textwidth]{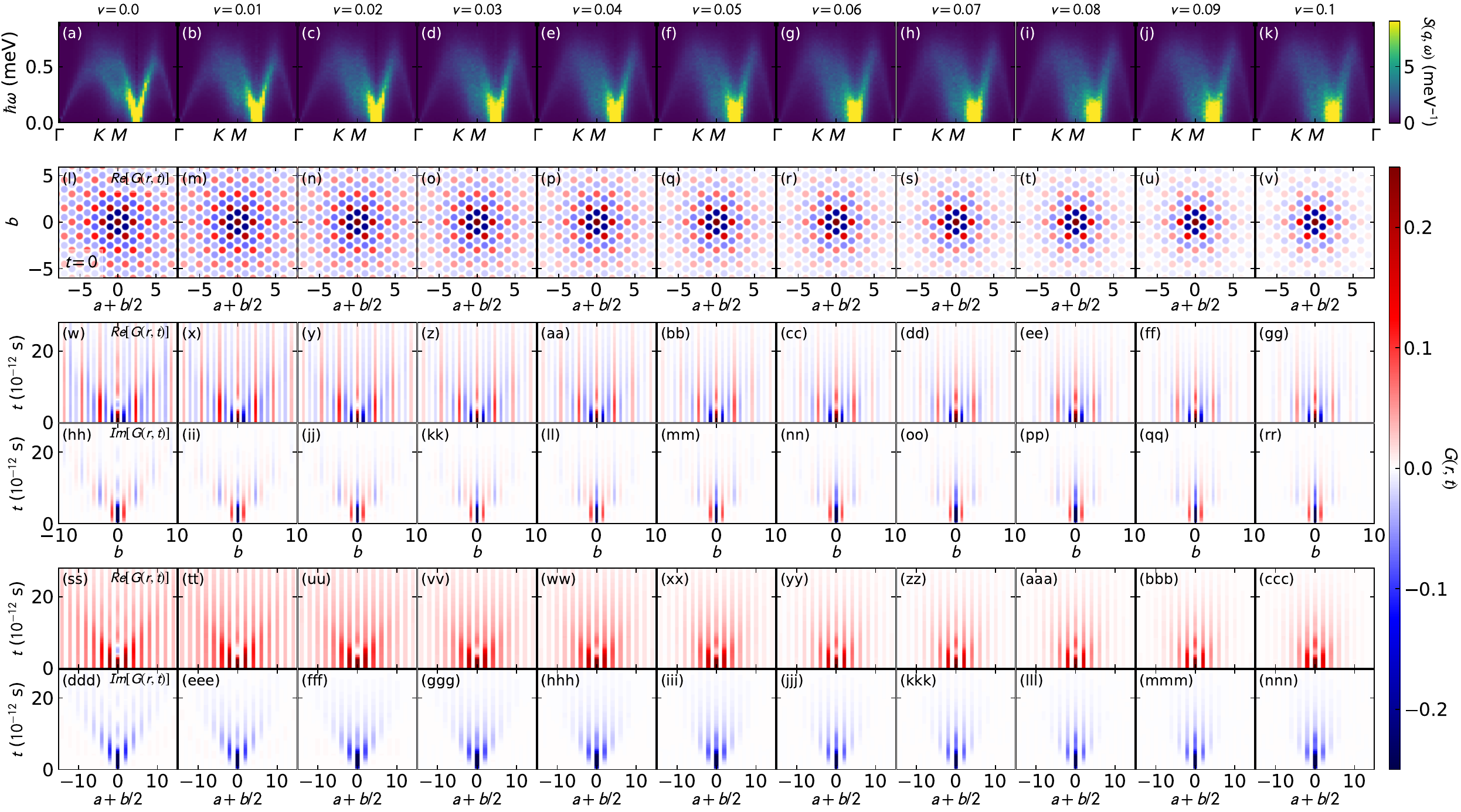} 
    \caption{Van Hove correlations for Landau-Lifshitz (LL) dynamics on the 2D triangular lattice with $J_2/J_1 = 0.044$ and $k_B T/ J_1 = 0.0625$ as a function of bond-vacancy probability $v$ (applied to both $J_1$ and $J_2$). 
    }
    \label{fig:LLD_vacancies}
\end{figure*}

\begin{figure*}
    \includegraphics[width=0.98\textwidth]{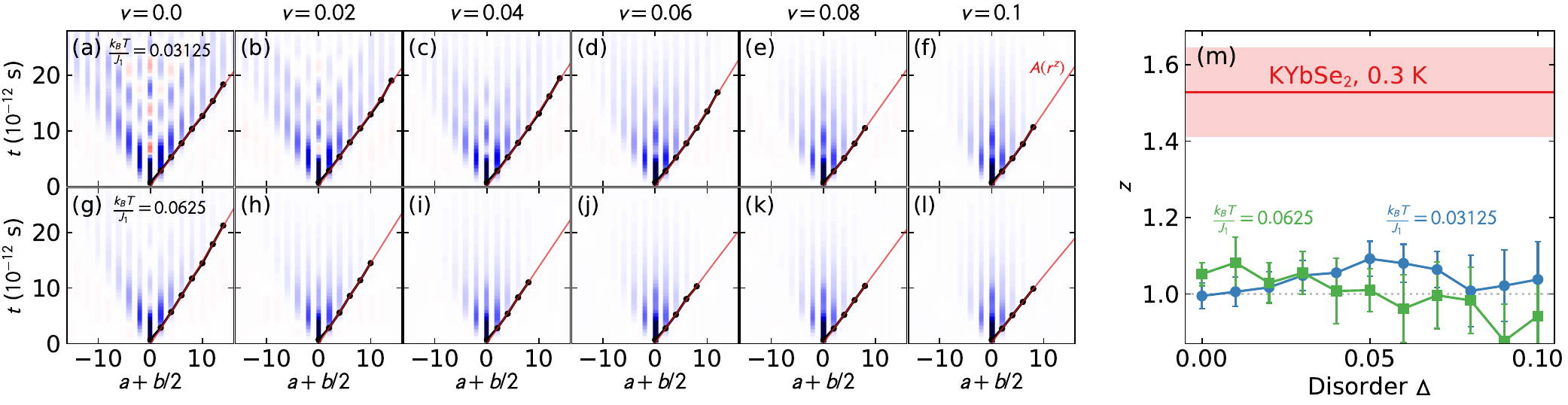} 
    \caption{Light cone onset along the second neighbor direction from LL simulations as a function of bond-vacancy disorder $v$. Panels (a)-(f) show the light cones at   $k_B T/ J_1 = 0.03125$, while panels (g)-(l) show the light cones at  $k_B T/ J_1 = 0.0625$. Black dots indicate the mid-point of the rise at different $r$, and the red line indicates a power law fit $A(r^z)$. 
    Panel (m) shows the fitted power law exponent for the different vacancy probability. This does not cause any significant nonlinearity. 
    }
    \label{fig:LLD_vacancies_LightCone}
\end{figure*}

Of course, random Gaussian disorder is not the only type of disorder possible: dilute vacancies and defects are another potential source of disorder in a real material. We simulate the effects of these in Figs. \ref{fig:LLD_vacancies} and \ref{fig:LLD_vacancies_LightCone}, where we simulate the effects of vacancies by randomly removing magnetic bonds from the lattice with a given probability. In this case, single-percent vacancies are enough to noticeably change the light-cone.  However, like the case of Gaussian exchange disorder, this does not cause a significant nonlinearity in the light-cone onset. 

These simulations show that, on the classical level, disorder is not a compelling explanation for the observed nonlinear light cone. The quantum situation may be much different, but is outside the scope of this study. 

\section{Relationship between real and imaginary $G(r,t)$}

The relation between real and imaginary $G(r,t)$ for a system in thermal equilibrium has a precise relationship determined by the fluctuation-dissipation theorem \cite{Sjolander1965theory_SI}. 
That is, if one knows the $Re[G(r,t)]$ and the temperature, one can derive  $Im[G(r,t)]$. 
Here we derive this relationship for $T=0$.

Let $G(r,t) = \mathcal{F}^{-1} (S(q,\omega))$, where $\mathcal{F}^{-1}(f(\omega)) = \int_{-\infty}^{\infty} d\omega \> e^{i \omega t} f(\omega)$ is an inverse Fourier transform. Neglecting position and momentum, we can write the real and imaginary parts as
\begin{equation}
    R(t) = \frac{1}{2} \left[ \mathcal{F}^{-1} (S(\omega)) + \mathcal{F}^{-1} (S(-\omega))^* \right]
\end{equation}
\begin{equation}
    I(t) = \frac{1}{2i} \left[ \mathcal{F}^{-1} (S(\omega)) - \mathcal{F}^{-1} (S(-\omega))^* \right].
\end{equation}
Now we note that by detailed balance,
\begin{equation}
    S(-\omega) = e^{-\beta \hbar \omega} S(\omega) 
    \label{eq:detailedbalance}
\end{equation}
where $\beta$ is the inverse temperature.  At $T=0$, $\beta = \infty$ and we can replace $e^{-\beta \hbar \omega}$ with the step function $u(\omega)$ which is 0 below $\omega = 0$ and 1 above $\omega = 0$. 
It is straightforward to show that 
\begin{equation}
    R(t) + iI(t) =  \int_{0}^{\infty} d\omega \>  e^{i \omega t} S(\omega)
\end{equation}
\begin{equation}
    R(t) - iI(t) =  \int_{0}^{\infty} d\omega \> e^{-i \omega t} S(\omega)
\end{equation}
from which we can write 
\begin{equation}
\int_{0}^{\infty} dt \sin(\omega t) R(t) = 
\int_{0}^{\infty} dt \cos(\omega t) I(t). 
\label{eq:SinTransform}
\end{equation} 
Thus to get $I(t)$ from $R(t)$, we take the inverse cosine transform of Eq. \ref{eq:SinTransform} (noting that $I(t)$ is even) and then take the sine transform. At nonzero temperature the equation is similar, but needs to keep the complex terms and apply the fluctuation dissipation theorem to relate the two sides of Eq. \ref{eq:SinTransform} \cite{Sjolander1965theory_SI}.

\end{document}